\documentclass[sigconf,nonacm]{acmart}
\pdfoutput=1 

\usepackage[normalem]{ulem}
\usepackage{booktabs}
\usepackage{xcolor}
\usepackage{listings}
\usepackage{verbatim}
\usepackage{graphicx}
\usepackage[htt]{hyphenat}
\usepackage{algorithm}
\usepackage{algorithmicx}
\usepackage{dblfloatfix} 
\usepackage{balance} 

\usepackage{cleveref}
\usepackage{textcomp}
\usepackage{xspace}
\let\OldTexttrademark\texttrademark
\renewcommand{\texttrademark}{\OldTexttrademark\xspace}%
\let\OldTextregistered\textregistered
\renewcommand{\textregistered}{\OldTextregistered\xspace}%
\usepackage{multicol}

\newcommand{\clwb}{\texttt{CLWB}}
\newcommand{\clflushopt}{\texttt{CLFLUSHOPT}}
\newcommand{\clflush}{\texttt{CLFLUSH}}
\newcommand{\movnti}{\texttt{movnti}}
\newcommand{\msq}{\texttt{MSQ}}
\newcommand{\dmsq}{\texttt{DurableMSQ}}
\newcommand{\liq}{\texttt{LinkedQ}}
\newcommand{\uq}{\texttt{Un\-lin\-kedQ}}
\newcommand{\olq}{\texttt{Opt\-LinkedQ}}
\newcommand{\ouq}{\texttt{Opt\-Un\-linkedQ}}
\newcommand{\iq}{\texttt{IzraelevitzQ}}
\newcommand{\nvtq}{\texttt{NVTraverseQ}}
\newcommand{\ofq}{\texttt{OneFileQ}}
\newcommand{\roq}{\texttt{Redo\-Opt\-Q}}
\newcommand{\cas}{\texttt{CAS}}
\newcommand{\nul}{\texttt{NULL}}
\newcommand{\vlp}{volatile linearization point}
\newcommand{\pp}{survival point}
\newcommand{\sfence}{\texttt{SFENCE}}
\newcommand{\persistent}{\texttt{Per\-sis\-tent}} 
\newcommand{\volatile}{\texttt{Volatile}}
\newcommand{\onll}{\texttt{ONLL}}

\usepackage[tt=false]{libertine}
\usepackage[scaled=0.82]{beramono}

\newcommand{\term}[1]{\textit{#1}}

\newtheorem{definition}{Definition}
\newtheorem{assumption}{Assumption}
\newtheorem{observation}{Observation}

\begin{document}

\title{Durable Queues: The Second Amendment}

\thanks{This work was supported by the United States - Israel BSF grant No. 2018655}

\author{Gal Sela}
\email{galy@cs.technion.ac.il}
\affiliation{
  \institution{Technion}
  \country{Israel}
}

\author{Erez Petrank}
\email{erez@cs.technion.ac.il}
\affiliation{
  \institution{Technion}
  \country{Israel}
}

\begin{abstract}
We consider durable data structures for non-volatile main memory, such as the new Intel Optane memory architecture. Substantial recent work has concentrated on making concurrent data structures durable with low overhead, by adding a minimal  number of blocking persist operations (i.e., flushes and fences). In this work we show that focusing on minimizing the number of persist instructions is important, but not enough. We show that access to flushed content is of high cost due to cache invalidation in current architectures. Given this finding, we present a design of the queue data structure that properly takes care of minimizing blocking persist operations as well as minimizing access to flushed content. The proposed design outperforms state-of-the-art durable queues.   

We start by providing a durable version of the Michael Scott queue (\msq{}). We amend \msq{} by adding a minimal number of persist instructions, fewer than in available durable queues, and meeting the theoretical lower bound on the number of blocking persist operations. We then proceed with a second amendment to this design, that eliminates accesses to flushed data. Evaluation shows that the second amendment yields substantial performance improvement, outperforming the state of the art and demonstrating the importance of reduced accesses to flushed content. The presented queues are durably linearizable and lock-free. Finally, we discuss the theoretical optimal number of accesses to flushed content.  

\end{abstract}




\settopmatter{printfolios=true}
\maketitle
The conference version of this paper is available at \cite{sela2021durablequeues}, and the code is publicly available at \url{https://github.com/galysela/DurableQueues}.

\section{Introduction}
\label{introduction-section}
Byte-addressable non-volatile memory combines DRAM's byte-ac\-ces\-si\-bility, with the durability and size of storage. 
Various technologies, such as resistive random access memory \cite{akinaga2010resistive}, phase-change memory \cite{raoux2008phase} and 3D XPoint \cite{3DX}, are expected to become available soon, with Intel/Micron 3D XPoint already available to consumers (under the brand name Optane). Non-volatile RAM (abbreviated as NVRAM) is expected to co-exist or replace DRAM in upcoming architectures, allowing program's modifications to its data structures survive system crashes. NVRAM platforms are expected to make a fundamental change in the design of the computing infrastructure including file systems, databases and other computations that process persistent data. 

While data stored in main memory will survive a crash, without further technological development, the caches and machine registers remain volatile, losing their content during a crash. This creates a consistency challenge, because writes may not reach the memory at the time and order the processor issues them.
When programs write data to memory, the CPU does not access the memory directly, but rather writes to the cache and the data only later gets flushed back to memory. Furthermore, the order in which cache lines get written back to the memory is unpredictable, as cache line evictions are triggered by local needs to make room for new cache content. 
This process may cause the state of the memory after a crash to become inconsistent, reflecting some modifications but missing others, impeding correct recovery.  

In order to make sure that the memory contains the required data for a potential crash and recovery, special instructions are used to force the flushing of cache lines from the cache to the memory. Asynchronous flush instructions initiate a cache line flush and let other instructions proceed while the data is being copied to memory. An additional synchronous fence (such as Intel's \sfence{} instruction) makes sure that the flushing becomes visible before any other memory instruction becomes visible to other threads. The fence instruction is blocking and costly and therefore durable algorithms have attempted to reduce the use of \sfence{} to achieve better performance. Cohen et al. \cite{cohen2018inherent} have shown that a durably linearizable \cite{izraelevitz2016linearizability} lock-free \cite{herlihy1991wait} object must use at least one fence instruction per update operation at worst case. They also presented a universal construction that achieves this bound, but their universal construction was intended as a proof of existence and no attempt was made to provide acceptable performance. 

The initial goal of this project was to optimize the performance of a durable FIFO queue. FIFO queues are used at the core of several existing persistent messaging systems (e.g., IBM MQ \cite{ibmmq}, Oracle Tuxedo MQ \cite{tuxedomq}, Rabbit MQ \cite{rabbitmq} and many more). Currently these queues are structured to suit the block-based interface of HDDs and SSDs. This design incurs costs like marshaling queue updates in streams, file system calls to persist message queues, etc., and so an adaptation to NVRAM platforms can bring a dramatic improvement to the queues performance and future use. 

Following previous work in this area, we focused on reducing the number of blocking persist operations. We started with the lock-free queue of Michael and Scott~\cite{michael1996simple} (denoted henceforth \msq{}), which was used in previous work~\cite{friedman2018persistent} due to its wide applicability to all architectures. We amended \msq\ in two different manners, obtaining two novel durably linearizable lock-free queue constructions with a minimal number of blocking persist operations: one blocking persist operation for any data structure modification operation. This meets the lower bound of Cohen et al.~\cite{cohen2018inherent}. These two optimal durable queue algorithms are the first contribution of this paper. 
 
One of these two algorithms, called \uq{}, is designed in the spirit of~\cite{zuriel2019efficient} to avoid persisting the underlying node links. In this algorithm, we allocate the nodes on designated areas, in which the recovery procedure can look for valid nodes of the queue. This requires a new persistent ordering mechanism that allows the recovery to determine the order of nodes in the queue without incurring a large overhead on the normal execution of the queue.
The second algorithm, denoted \liq{}, does persist the underlying node links. It reduces the number of fences by using a validity scheme to inform the recovery algorithm which nodes are adequate for recovery. It also adds a backward link to the queue nodes, for enabling to efficiently assist persisting concurrent operations.

We implemented these two algorithms on a platform with an Intel Cascade Lake processor and an Intel Optane NVRAM. Surprisingly, the new algorithms did not show a clear improvement over the state-of-the-art durable queue of Friedman et al. \cite{friedman2018persistent} although Friedman's queue executes more blocking persist operations during the execution. Further investigations raised an interesting problem. Our queues frequently access flushed cache lines, and these accesses significantly deteriorated performance. It tur\-ned out that Intel flush instructions, which flush a cache line to the NVRAM, cause the flushed cache line to be invalidated in the cache, so that subsequent accesses yield cache misses and re-read the data from memory. (We tried various instructions including the most advanced \clwb{} instruction, but they all had the same performance degradation effect). The resulting additional loads from memory are significantly more costly on NVRAM than on DRAM, due to the high NVRAM read latency. 
While the recently-launched Intel Ice Lake processors with Optane persistent memory 200 series may provide flush instructions that do not invalidate the flushed cache lines, existing NVRAM architectures with Cascade Lake processors do not seem to support such instructions. Our impression is corroborated in the findings of \cite{wen2020montage,kalia2020challenges,van2019persistent,clwb-blog-post,bu2021revisiting,friedman2020nvtraverse,friedman2021mirror}. Existing (costly) architectures will probably remain in use for years to come and one needs to use algorithmic modifications to obtain improved performance on such machines.  

We amended the two algorithms further, obtaining algorithms that avoid accessing flushed locations. While changing the algorithms, we made sure that their original advantage of a single fence per update operation is maintained. 
An evaluation of this second amendment demonstrates a significant performance improvement, which confirms the high cost of accessing flushed content on these platforms. 

The second contribution of this paper is a guideline for designing durable data structures and algorithms for NVRAM. In addition to the well-known guideline to minimize blocking persist operations, we recommend designing algorithms with reduced access to recently flushed cache lines\footnote{We consider only explicitly flushed cache lines. There are additional implicit flushes, e.g., when the system evacuates cache lines to make space for new lines that need to be loaded to the cache. Such implicit flushes are hard to predict and this guideline does not attempt to consider them.}. This guideline is relevant for platforms that invalidate cache lines when flushing their content to the memory, and the purpose is to avoid the cost of fetching data from the memory after it is evicted from the cache. This guideline is especially important in light of the high read latency of available NVRAM (see measurements in \cite{van2019persistent,yang2020empirical}). 

Our third contribution is the design of durable lock-free queues with significantly improved performance for the new Intel Optane architecture. We present \ouq{} and \olq{}, obtained by amending \uq{} and \liq{} respectively according to the new guideline. \ouq{} and \olq{} are the best performing lock-free durable queues available today. We compare the performance of \ouq{} and \olq{} against state-of-the-art durable queues and against \uq{} and \liq{} themselves, which use minimal blocking persist operations but do not consider the new guideline and do not reduce access to flushed cache lines. While \ouq{} and \olq{} outperform all other queues on nearly all thread counts and workloads, we believe \uq{} and \liq{} are still interesting to present. This is because for potential more advanced platforms that might provide flushing without cache invalidation, \uq{} and \liq{} may turn out best. 

From a theoretical standpoint, it is interesting to note that \ouq{} and \olq{} yield the best possible design characteristics for durability. Following our guideline above, they make zero accesses to content that was previously (explicitly) flushed, while they also meet the lower bound shown by Cohen et al.~\cite{cohen2018inherent}, executing only a single blocking persist operation per data structure update operation. Interestingly, while these theoretical characteristics are the best possible, they are also obtainable for any object. This follows from the universal construction of~\cite{cohen2018inherent}. While Cohen's universal construction of lock-free durably linearizable data structures is not practical, it has the above-mentioned characteristics (a single blocking persist instruction per update operation and no access to flushed content) and it is applicable to any object. 

The rest of the paper is organized as follows. In \Cref{model-section} we elaborate on the model and the general upper bound on the design parameters. In \Cref{preliminaries-section} we recall the definitions of durable linearizability and lock-freedom as well as \msq{}, the basic queue algorithm that we extend in our constructions. 
We discuss related work in \Cref{related-work-section}.
In \Cref{first-amendment-section} we provide an overview of the main ideas in the first amendment to \msq: minimizing blocking persist operations, which produces \uq{} and \liq{}. 
In \Cref{principles-to-reduce-accesses-after-flush-section} we describe the second amendment to the two algorithms, adhering to the guideline of reducing access to flushed data, which results in the optimal queues \ouq{} and \olq{}.
The details of the \uq{} algorithm are provided in \Cref{first-amendment-section}, while further details of the rest of our queues are deferred to \Cref{liq-details,ouq-details,olq-details}.
We argue about the durable linearizability and lock-freedom of our queues in \Cref{durable-linearizability-sketch,Lock-Freedom-sketch}.
The memory management scheme applied in our queues is described in \Cref{mm-section}, and the performance of all queue algorithms is evaluated in \Cref{performance-section}. We conclude in \Cref{conclusion-section}.

\section{Model}
\label{model-section}

In the persistent memory model, there are two levels of memory -- volatile (registers, caches) and persistent (NVRAM). Values in the cache may be written back to the persistent memory implicitly by a cache eviction, or explicitly by flush instructions.
We adopt the failure model of Izraelevitz et al. \cite{izraelevitz2016linearizability} for crashes, which considers full-system crashes in which all processes fail together. 
The state of the volatile memory is lost in a crash, but the state of the persistent memory remains unaffected.
After a crash, new threads are created and proceed with the computation.
Each data structure may provide a recovery procedure to be invoked after the crash for restoring a consistent state of the object from its preserved state in the NVRAM.
Our data structures apply a complete recovery before continuing with any new operation.

To maintain correctness in the presence of crashes, one has to ensure that necessary writes propagate from the cache to the persistent memory.
To ensure a written value becomes persistent (after being written to the cache), one may issue a flush instruction and block until it completes. A flush instruction receives a memory address and flushes the content of the cache line containing this address to the persistent memory. Some flush instructions are asynchronous, enabling issuing multiple flushes concurrently as an optimization. Subsequently, a store fence instruction, denoted \sfence{} (like the instruction name on Intel), may be placed to ensure completion of all previous asynchronous flushes. Throughout the paper, when mentioning a {\em persisting} of a location, we refer to an asynchronous flush of its address accompanied by an \sfence{} to ensure that the data in this location has been written to the NVRAM.

Intel flush instructions (such as the synchronous \clflush{} and the asynchronous \clflushopt{} and \clwb{}) take a memory location and write back the cache line containing it to the memory, if this line consists of modified data. 
According to the Intel architectures software developer's manual \cite{IntelSDM}, \clflush{} and \clflushopt{} do not only write the cache line to the memory, but rather also invalidate it. Regarding \clwb{}, the Intel manual states that it {\em may} retain the line in the cache. However, on the Second Generation Intel Xeon Scalable Cascade Lake processor we use, \clwb{} seems to invalidate the cache line like \clflushopt{} does: replacing \clwb{} with \clflushopt{} in all the data structures we measured yielded similar performance. This is also noted by others \cite[e.g.][]{wen2020montage,kalia2020challenges,van2019persistent,clwb-blog-post,bu2021revisiting,friedman2020nvtraverse,friedman2021mirror}.
The recently-launched Third Generation Intel Xeon Scalable Ice Lake processors with Optane persistent memory 200 series may implement \clwb{} retaining lines in the cache, but NVRAM platforms currently in the market do not seem to support flushes without cache invalidation. 
Existing architectures will probably remain in use for years to come.
Therefore, designers of efficient durable algorithms should take into consideration the cost of accessing a memory location after it was flushed and evicted from the cache. 

To eliminate some of the costly persisting occurrences, we rely on the following assumption, which is based on the cache line granularity of write backs to memory.
The assumption is mentioned in the SNIA NVM programming model \cite[Section 10.1.1]{SNIA}, adopted by Intel for working with persistent memory (as stated in Intel's formal persistent memory programming book \cite{scargall2020programming}), and is confirmed by Intel Senior Principal Engineer Andy Rudoff in online informal discussions \cite[e.g.][]{google-group-1-confirming-atomic-cache-line-writeback,google-group-2-confirming-atomic-cache-line-writeback,stackoverflow-confirming-atomic-cache-line-writeback}.
This assumption was also previously made in \cite[Footnote 16]{chakrabarti2014atlas} and \cite[Assumption 2]{cohen2017efficient}.
\begin{assumption} \label{release-fence-adequate-between-writes-to-same-cache-line}
A cache line is evicted atomically to memory, thus, the order of multiple writes to the same cache line is preserved in memory. 
In other words, the content of a cache line in the memory reflects a prefix of the stores to that cache line. 
\end{assumption}

As the order of writes to the same cache line is preserved in NVRAM\footnote{Writes to the cache are not guaranteed to occur in program order, due to compiler optimizations, but program order can be enforced by placing inexpensive release fences (that prevent compiler optimizations, thus, ordering writes to cache). We placed release fences in our implementation where required, and we do not further mention them here.}, placing a flush plus \sfence{} between them to ensure their persistence order (which is required for writes to different cache lines) is redundant.

In addition to a flush, another useful instruction for our algorithms is an instruction that writes back data directly to the memory without touching or polluting the cache (like \movnti{}). Such asynchronous instructions require an accompanying \sfence{} to ensure their completion.

\subsection{Upper Bound on Accesses after a Flush}
\label{upper-bound-subsection}
Due to cache invalidation after a flush, we recommend designing algorithms that minimize accesses to flushed content. This comes in addition to designing algorithms that minimize blocking flushes. In fact, we claim that it is possible to implement any object with a deterministic sequential specification in a durably linearizable lock-free way using the minimum possible number of fences (one per update operation and zero per read-only operation, as proved by \cite{cohen2018inherent}) while at the same time performing zero accesses to (explicitly) flushed cache lines. 

To prove our claim, we leverage the universal construction of \cite{cohen2018inherent}, called \onll{}. \onll{} consists of two main components. The first is a shared execution trace, containing a mark indicating the trace's prefix guaranteed to be persistent. This prefix represents the current state of the object. The execution trace is not used during recovery, thus also not persisted to memory. The second component is local per-thread persistent logs (adopted from \cite{cohen2017efficient}), that will be read during recovery.
An update operation first appends a record representing it to the execution trace, then appends a copy of the trace's suffix that is not yet guaranteed to be persistent to its local log and persists it, and finally marks the trace's prefix up to the current operation as persistent. A read-only operation calculates its response based on the current state of the object, represented by the trace's marked prefix. 

\cite{cohen2018inherent} proves that \onll{} obtains the minimum possible number of fences. We suggest the following slight modification to \onll{}: align log entries to cache lines, so that no two entries will share a cache line. 
By applying this modification, \onll{} still obtains minimum fences, while also performing no access to flushed memory. This is because only data in the local per-thread persistent logs is explicitly flushed, and these logs' cache lines are not accessed after their flush: they are read only during recovery, and not written by following log appends -- which write to following cache lines thanks to our modification.
\section{Preliminaries for the Durable Queues}
\label{preliminaries-section}

\subsection{MS-Queue}
Our persistent queue algorithms extend the widely used \msq{} (the Michael and Scott queue \cite{michael1996simple}), a well-performing concurrent queue adequate for general hardware, included as part of the Java\texttrademark Concurrency Package \cite{lea2009java}. This is a (non-persistent) lock-free FIFO queue, which supports enqueue and dequeue operations. It implements the queue as a singly-linked list with head and tail pointers. Nodes in the list have two fields: a value and a next pointer. The head points to the first node of the list, which functions as a dummy node. Subsequent nodes, after the dummy and until the node whose $next$ pointer's value is \nul{}, contain the queue's items. The queue is initialized to an empty queue as a list that contains a single (dummy) node, to which both the head and tail point. 

A dequeue operation checks if $next$ of the obtained head is \nul{} (meaning the queue is empty). If so, this is a {\em failing dequeue} that returns without extracting an item from the queue. Otherwise, an attempt is made to update the head to point to its successive node in the list, using a \cas{}, and on failure the dequeue operation starts over. A dequeue that succeeds to perform a \cas{} that advances the queue's head is denoted a {\em successful dequeue}.

Enqueuing requires two \cas{} operations. Initially, a node with the item to enqueue is created. Then, an attempt to set $tail\text{->}next$ to the address of the new node is made using a first \cas{}. The \cas{} fails if the value of $tail\text{->}next$ is not \nul{} in that moment. In such a case, an attempt to advance $tail$ to the current value of $tail\text{->}next$ is made using a \cas{}, to help an obstructing enqueue operation complete. Then, a new attempt to perform the first \cas{} starts. After the first \cas{} succeeds, a second \cas{} is applied to update $tail$ to point to the new node.

\subsection{Linearizability and Durable Linearizability}
Defining correctness for durable executions in the presence of both concurrency and NVRAM is not a trivial task. 
In this work, following recent work in this domain, we adopt durable linearizability \cite{izraelevitz2016linearizability} described below as a correctness criterion. Nevertheless, it is easy to verify that our proposed queues satisfy also other correctness criteria, like strict linearizability \cite{aguilera2003strict}, persistent atomicity \cite{guerraoui2004robust} and recoverable linearizability \cite{berryhill2015robust}.

We recall some basic terminology. An \term{operation} consists of two events -- \term{invocation} and \term{response}. 
An execution of a concurrent system in the full-system-crash model may be modeled by a finite sequence of events of three types: invocation events and response events, each tied to specific process and object, and system crash events (which are not tied to a specific process or object). Such sequence is denoted a \term{history}.
An operation in a given history is \term{pending} if the history contains only its invocation event and not its response event. 
We refer to an operation for which the history contains also the response as \term{completed}.
Each object has a \term{sequential specification}, which describes its behavior in sequential executions, where operations do not overlap.

A history without crash events is considered \term{linearizable} \cite{herlihy1990linearizability,sela2021linearizability} if each completed operation appears to take effect at once, between its invocation and its response events, in a way that satisfies the sequential specification of the objects. Each pending operation is required to either take effect at once after its invocation in a way that satisfies the sequential specification of the objects, or not take effect at all.
A history in the full-system-crash model (i.e., a history that might contain crashes in which all processes fail together and there is no subsequent thread reuse) is considered \term{durably linearizable} \cite{izraelevitz2016linearizability} if the history with the crashes omitted is linearizable.

\subsection{Lock-Freedom}
\label{preliminaries:Lock-Freedom}
A concurrent object implementation is \term{lock-free} \cite{herlihy1991wait} if each time a thread executes an operation on the object, some thread (not necessarily the same one) completes an operation on the object within a finite number of steps. 
We extend the definition to executions with crashes, and define
a concurrent object implementation to be lock-free in the presence of crashes if each time a thread executes an operation on the object,
and there are no interrupting crash events since the operation's invocation,
some thread (not necessarily the same one) completes an operation on the object within a finite number of steps.
This definition is equivalent to the one brought in \cite{zuriel2019efficient}, which considers crashes as progress, as if a crash is one of the operations on the data structure.
Lock-freedom guarantees system-wide progress. Our implementations are lock-free.
\section{Related Work}
\label{related-work-section}

There has been a large body of work by multiple communities that provides algorithms for NVRAM. 
Several libraries for persistent transactional access to objects in NVRAM have been proposed \cite{coburn2011nv,correia2018romulus,kolli2016high,marathe2018persistent,ramalhete2019onefile,pmem,volos2011mnemosyne,wu2019architecture,zardoshti2019optimizing}, but persistent transactions require heavy-duty logging mechanisms, and thus do not yield highly efficient solutions, and are not competitive with ad-hoc constructions such as ours. 
\cite{memaripour2020pronto} presents an NVRAM library taking another logging-based approach.
Izraelevitz et al.~\cite{izraelevitz2016linearizability} suggested to automatically make concurrent objects durably linearizable by adding a flush and a fence after each access to global memory (a read or a write). This transformation yields a durable variant of any existing lock-free data structure, but the resulting implementations are typically inefficient. The first ad-hoc efficient lock-free durable data structure was the queue presented by Friedman et al.~\cite{friedman2018persistent}, with a substantial reduction of the number of fences executed with each operation over the general construction of Izraelevitz. Subsequently, David et al.~\cite{david2018log} presented lock-free durable set implementations (including a linked list, a skip list and a hash map). Zuriel et al.~\cite{zuriel2019efficient} improved over that construction and presented a set with a single \sfence{} per update operation, thus meeting the lower bound of Cohen et al.  \cite{cohen2018inherent} and also obtaining much better performance.   
Raad et al. \cite{raad2020persistency} implemented a persistent FIFO queue to demonstrate the application of their suggested hardware model, but did not aim for optimized performance (e.g., they do not track the tail pointer, thus significantly slowing down enqueues).
\section{First Amendment: Queues with minimum fences}
\label{first-amendment-section}
The current literature offers a fast queue with several fences \cite{friedman2018persistent} on the one hand, and a universal construction for all data structures with a single fence (per update operation) which is extremely inefficient~\cite{cohen2018inherent} on the other hand. A question that naturally arises is whether it is possible to reduce the number of fences to the minimum possible number, and at the same time be able to use this reduced blocking to obtain a better performing queue. 
In this section we provide two durably linearizable lock-free queues, \uq{} and \liq{}, that meet the theoretical lower bound on the number of fences (a single \sfence{} per operation). 

\subsection{\uq{}}
As its name implies, \uq{} does not rely on links between nodes for restoring the queue after a crash and therefore does not persist them, similarly to the basic idea in~\cite{zuriel2019efficient}. It keeps all information required for recovery in the nodes themselves, which are located in designated areas. Upon a crash, the recovery procedure checks these nodes to decide which ones are valid and belong to the resurrected queue. 
The links are still used to expedite operations on the queue when no crash occurs, but they are not required to reconstruct the queue after a crash. Care is taken to persist the queue order in the nodes to allow proper recovery.

\uq{} places {\em index} and {\em linked} fields in each node to enable the recovery to identify which nodes in the designated areas should be restored and in what order.
The {\em index} field states the node's index in the queue (according to enqueue order). 
Overflow can be handled, but for now we allocate 64 bits for the {\em index} field and assume that it does not overflow (while humans are still around).
The {\em linked} field marks nodes that have been added to the queue. After an enqueuer succeeds to link a node to the queue, it sets its {\em linked} flag, and then persists the node content. 
The recovery procedure resurrects nodes that are marked {\em linked} and have an index larger than the head, and arranges them in the order induced by their indices.
After advancing the head, a dequeuer persists the new head's index, to indicate to the recovery that all nodes up to this one are dequeued.
This scheme forms a consecutive prefix of dequeued nodes -- all those that the head has persistently passed, thus satisfying the FIFO order requirement.

The simple scheme described so far involves several races which should be resolved. 
One race stems from the fact that the order in which enqueue operations  complete does not necessarily match the linking order of their nodes. For example, it is possible that the enqueue of the fourth node in the queue has completed before the enqueue of the third node in the queue completed. Hence, it might be that the fourth node is marked {\em linked} while the third node is not. One consequence of this race is that the indices of valid linked nodes that the recovery identifies do not always form a sequence of consecutive integers. 
Even worse, a dequeue operation might point the head at a node inserted by a concurrent enqueue, whose content is not yet flushed and therefore contains a stale index that may confuse the recovery. 

Next, we elaborate on the implementation of \uq{}, including describing how it resolves the above-mentioned issues.
The \uq{} algorithm is presented in \Cref{fig:uq-impl}. A description of its operations follows.

\begin{figure}
\caption{\uq{} implementation}\label{fig:uq-impl}
\vspace{0.5\baselineskip}
\begin{minipage}{.49\textwidth}
\begin{lstlisting}
class Node
    Item* item
    atomic<Node*> next
    bool linked
    int index
\end{lstlisting}
\begin{lstlisting}
@\underline{Item* Dequeue()}@
    while (true)
        head = Head
        headNext = head.ptr->next@\label{uq-read-head-next}@
        if (headNext == NULL)@\label{uq-deq-fails-start}@
            FLUSH(&Head.index); SFENCE@\label{uq-persist-failing-deq}@
            return NULL@\label{uq-deq-fails-end}@
        if (CAS(&Head, head, @$\langle$@headNext, headNext->index@$\rangle$@)@\label{uq-deq-cas}@
            dequeuedItem = headNext->item
            FLUSH(&Head.index); SFENCE@\label{uq-persist-successful-deq}@
            if (nodeToRetire[tid]) // It equals NULL in the first successful dequeue
                retire(nodeToRetire[tid])@\label{uq-deq-retire}@
            nodeToRetire[tid] = head.ptr@\label{uq-save-node-to-reclaim}@
            return dequeuedItem
\end{lstlisting}
\end{minipage}\hfill
\vspace{0.1\baselineskip}
\begin{minipage}{.49\textwidth}
\begin{lstlisting}
@\underline{Enqueue(item)}@
    newNode = allocNode()@\label{uq-alloc-node}@
    newNode->item = item
    newNode->next = NULL@\label{uq-enq-set-next-null}@
    newNode->linked = false@\label{uq-unset-linked}@
    while (true)
        tail = Tail
        if (tail->next == NULL)
            newNode->index = tail->index + 1@\label{uq-set-index}@
            if (CAS(&tail->next, NULL, newNode))@\label{uq-link}@
                newNode->linked = true@\label{uq-set-linked}@
                FLUSH(newNode); SFENCE@\label{uq-persist-enq}@
                CAS(&Tail, tail, newNode)@\label{uq-advance-tail}@
                break
        CAS(&Tail, tail, tail->next)@\label{uq-enq-assisting-cas-tail}@
\end{lstlisting}
\end{minipage}
\end{figure}

\subsubsection{The Enqueue Operation}\label{uq-enq-description}

The enqueue operation first allocates a node and initializes its data (\Crefrange{uq-alloc-node}{uq-enq-set-next-null}).
It then unsets {\em linked} (\Cref{uq-unset-linked}), 
sets the {\em index} of the new node to be the index of the last node plus one (\Cref{uq-set-index}),
and attempts to link the node to the queue (\Cref{uq-link}).
The reason {\em linked} is unset before {\em index} is updated, is that when the node is allocated, its {\em linked} flag might be set; thus, assigning the new node a relevant index in this state might erroneously cause the recovery to restore the node even though it is not yet linked to the queue.

After succeeding to link the node, the enqueuer sets its {\em linked} flag (\Cref{uq-set-linked}), to signal to the recovery (that would run if a crash occurs) that the node should be restored.
The described order of writes to the node fields guarantees, based on \Cref{release-fence-adequate-between-writes-to-same-cache-line}\footnote{Applying \Cref{release-fence-adequate-between-writes-to-same-cache-line} requires that the whole node resides on a single cache line, which is typically the case, and it also holds for the queues implemented in this paper. The method of~\cite{cohen2017efficient} can be used to generalize the algorithms to nodes that span multiple cache lines without adding fence operations.}, that a node will be restored by the recovery only if it is successfully linked.        
Finally, the enqueuer persists the node and advances the queue's tail to point to the new node (\Crefrange{uq-persist-enq}{uq-advance-tail}).
If a concurrent enqueue operation interferes, the enqueuer attempts to assist the other enqueue to advance the tail to point to its node (\Cref{uq-enq-assisting-cas-tail}), before starting a new attempt to enqueue its own item.

We note that the recovery procedure might restore a suffix of enqueues with nonconsecutive indices. This happens only if several enqueues are running when a crash occurs: an enqueue that linked e.g. the fourth node in the queue might have set its {\em linked} flag and persisted it before the crash, while an enqueue that linked the third node in the queue has not. Discarding pending enqueue operations which have not set and persisted the {\em linked} flag is correct due to the following observation:
\begin{observation}\label{observation:nonconsecutive-enqueues}
Durable linearizability allows pending operations to not be linearized.
Therefore, the recovery may discard pending enqueue operations, which might result in a suffix of enqueued nodes with nonconsecutive indices.
\end{observation}

\subsubsection{The Dequeue Operation}\label{uq-deq-description}
If a dequeue operation encounters an empty queue, it returns \nul{}.
Otherwise, it attempts to advance the head by one node, and on success -- it returns the oldest item to the caller. On failure it retries the whole scheme.

To signal to the recovery procedure that it should ignore the dequeued node, a successful dequeue operation ensures that the head's index is persistently increased to a value bigger than or equal to its dequeued node's index. 
Persisting the new head's index is intended to indicate to the recovery not only that this node is dequeued, but also that all nodes up to this one are dequeued, and a failing dequeue also needs to persist the head's index before returning in order to persist the previous dequeues that emptied the queue.
This is obligatory due to the following observation:
\begin{observation}\label{observation:consecutive-dequeues}
The recovery must restore a consecutive prefix of dequeued nodes, to satisfy the FIFO order requirement.
\end{observation}
The recovery achieves this by interpreting nodes with index smaller than or equal to the head's index as dequeued.

A successful dequeue is responsible to reclaiming the node that was the head during the previous dequeue that this thread executed. This node to be retired is kept in a {\em nodeToRetire} array, consisting of a cell per thread. Its cells do not share cache lines to avoid false sharing. Each thread may access its cell using its thread ID as an index. 

Next, we explain how dequeuers ensure that the correct head's index is restored by the recovery.
If we let a dequeuer persist the head's address, and let the recovery determine the head's index to be the {\em index} in the node pointed to by this head (as appears in NVRAM in the crash moment), then the recovery might erroneously restore a stale (smaller) head's index value, and discard completed dequeues. This could happen if the enqueuer of the node pointed to by the head has linked the node but was interrupted by the crash before persisting the node's data.
Therefore, \uq{} takes a different approach to determine the head's index in recovery.

\uq\ makes the head hold not only a pointer to the dummy node, but also its index. They are held side-by-side and updated together atomically using a double-width \cas{}. 
A dequeuer starts by performing a double-width \cas{} (\Cref{uq-deq-cas}) that advances the head's pointer and increments the head's index. Next, the dequeuer persists the index placed in the head (\Cref{uq-persist-successful-deq}). A failing dequeue assists persisting the head's index too (\Cref{uq-persist-failing-deq}).\footnote{We particularly specify the head's index as the flushed value in order to stress that this is the data required in recovery, but its flush clearly writes the whole containing cache line to the memory.} The recovery procedure restores the head's index from the value kept in the queue's head, rather than from the possibly stale value in the node pointed to by the head.
This prevents discarding a completed dequeue: persisting the head's index after incrementing it to the index of the dequeued node, makes the recovery procedure ignore the dequeued node. 

The use of a double \cas{} can be eliminated (if the platform does not support it) by taking an alternative approach: Each thread could maintain a local index. After each time it advances the queue's head, it would update the local index with the value of the new head's index and persist it. The recovery would then restore the head's index as the maximum across these local indices.
The alternative handling of the head's persistence described here, is actually required and applied in the second amendment of \msq{} (see \Cref{principles-to-reduce-accesses-after-flush-section}).

\subsubsection{Recovery}
\label{uq-recovery-subsection}
The recovery procedure of \uq{} resurrects nodes in the designated areas that are marked {\em linked} and have an {\em index} bigger than the head's index. It then sets their links to form a linked list that holds the queue nodes in the order induced by their indices.
This is implemented as follows. 

The head's index is not modified. A dummy node is allocated and assigned an index that matches the head's index. The head's pointer is set to point at this dummy node. 
Next, the recovery scans the designated areas and makes a list of recovered nodes, which are those with a set {\em linked} flag and an {\em index} larger than the head's index. 
All other nodes are reclaimed. The recovered nodes are then sorted and their {\em next} pointers are set accordingly to create the queue. Finally, the queue's tail is set to point to the last node in the queue. 

We note that free nodes (owned by the memory manager) in the designated areas are appropriately ignored by the recovery:
When the memory manager allocates a new designated area for nodes from the operating system, it zeros its content, to make all nodes consist of a zeroed index, and then persists it in NVRAM (by placing asynchronous flushes of the whole area accompanied by a single \sfence{}). 
If the number of required nodes is unknown in advance, each time a designated area is depleted, the memory manager may allocate a new area from the operating system and initialize it in a similar manner using a single \sfence{}. 
The zeroed indices guarantee that the unused nodes owned by the memory manager are ignored by the recovery.
In addition to these not-yet-allocated nodes, nodes reclaimed by dequeuers are also ignored by the recovery thanks to their index value, as dequeue operations return nodes to the memory manager only after the head's index persistently equals to the index of a subsequent node. 
Finally, nodes reclaimed by a previous recovery process are ignored thanks to either their {\em index} or their unset {\em linked}.

\subsection{\liq{}}
\label{liq-overview-subsection}
\liq{} also performs a single fence in each operation, but using a completely different approach.
Here, we provide an overview of \liq{}, and the full details appear in \Cref{liq-details}.

The first idea \liq{} employs is to make the recovery procedure able to deal with nodes whose data has not 
been persisted. This allows linking nodes to the queue without blocking to persist their data beforehand, thus avoiding one of the fences of the queue in \cite{friedman2018persistent}. 
To enable this, \liq{} presents a mechanism that identifies nodes with stale data:
a designated {\em initialized} flag in each node signifies whether the content of the node is guaranteed to be valid. We maintain the invariant that if the node's data is not initialized in NVRAM, then its {\em initialized} flag is unset in NVRAM. 
To achieve this, \liq{}'s enqueue operation initializes the node in two steps: first, it initializes the node content, and then it sets the {\em initialized} flag. No \sfence{} is issued during this execution, as \Cref{release-fence-adequate-between-writes-to-same-cache-line} guarantees that the order of writes to the same cache line is not reversed. 

For this scheme to work, we need to make sure that when a node is allocated, its {\em initialized} flag is unset. This can be easily done with an extra fence at allocation time, but would yield two fences per enqueue operation. We manage to avoid this fence by postponing the return of dequeued nodes to the memory manager. 
Think first of a simplified version that lets each thread accumulate $k$ nodes it removed from the queue. After each $k^{th}$ successful dequeue, before returning the $k$ nodes to the memory manager, the thread clears their {\em initialized} flags, issues an (asynchronous) flush for each of the flags, and then a single blocking fence before letting the memory manager reclaim these objects. Such a simplified algorithm would execute $1+1/k$ fences per successful dequeue operation, not perfectly meeting the desired theoretical lower bound of a single fence. 
To reduce the number of fences to one, we take a more complex approach: After removing a node $N$ from the queue, its dequeueing thread $T$ clears $N$'s {\em initialized} flag and records $N$'s address for later. Instead of placing an additional fence every $k$ dequeues, $T$ will piggyback on the fence which its next successful dequeue anyhow performs: $T$ will flush $N$'s {\em initialized} flag before this fence, and return $N$ to the memory manager after this fence.
Such piggybacking on a fence of a later operation by the same thread makes sure that {\em initialized} flags are properly reset in memory before their nodes are reused, without incurring additional fences. 

The recovery procedure resurrects all nodes reachable from the head through a path of consecutive nodes with the {\em initialized} flag set. 
It remains to ensure that completed enqueue operations are visible to the recovery procedure, even though previous nodes\footnote{We think of the nodes as ordered by the underlying linked list of the queue. This order enables the terms {\em previous, preceding, subsequent, consecutive,} etc.} in the queue may have been enqueued by operations that have not yet completed.  
Before an enqueue operation completes, \liq{} makes sure that 
all data on nodes from the head to the enqueued node is written back to the NVRAM. This guarantees that the recovery will reach the new node in its traversal from the head.  
Naive\-ly, before an enqueue operation completes, the enqueuer could traverse all nodes from the head until the new node, flush their contents, and then issue a single fence. This persists all relevant nodes but at a very high cost. To make this process efficient, we add a backward edge to the underlying linked list, and walk backwards persisting only nodes that might have not yet been persisted. We attempt to minimize the length of walk as much as possible. 
The full details of \liq{} appear in  \Cref{liq-details}.

\section{Second Amendment: Queues with no post-flush access}
\label{principles-to-reduce-accesses-after-flush-section}

It turns out in the evaluation that reducing the number of fences is not enough to obtain high performance, and one should further improve the algorithms by reducing accesses to flushed data. In this section we describe further transformations of \uq{} and \liq{} into the new algorithms \ouq{} and \olq{} respectively which do not access flushed locations, while still executing the minimal possible number of blocking fences per operation. Evaluation will show that the obtained algorithms yield excellent performance in current architectures. 
These algorithms are the fastest available persistent queues today, but we believe that \uq{} and \liq{} are of value on their own. This is because future architectures may provide flushes that do not invalidate cache lines. In such architectures \uq{} and \liq{} are expected to perform well thanks to using the minimal number of fence instructions. However, we cannot evaluate this performance prediction on the platform we currently possess. 

\subsection{\ouq{}} \label{ouq-details-description}
We provide an overview of \ouq{} here, and detail its pseudocode in \Cref{ouq-details}.
We start with looking at what data is flushed in the \uq{} algorithm, for use in a recovery. 
\uq{} flush\-es the global head index, plus, the {\em index}, {\em item} and {\em linked} fields for each node in the underlying linked list. 
All of these values except for the {\em linked} field are later accessed. We eliminate these accesses using algorithmic modifications, amending \uq{} to become \ouq{}. 

First, we switch the global head index with a per-thread head index, holding the value that the head index had during the last dequeue by the thread. In \ouq{} the head pointer is a pointer only (with no adjacent index). Instead of persisting the global head index in the end of every dequeue operation as \uq{} does, a dequeuer of \ouq{} copies the index value of the node pointed to by the head pointer to its local head index and persists it. In a recovery, the head index is set to the maximal index among the local head indices of all threads.
Note that in this description we write to the local head index after persisting it. We eliminate this access in Section~\ref{sec-opti} below. 

The {\em index} and {\em item} fields of a node in \uq{} are written by the node's enqueuer, and then (after the node is linked to the queue) -- flushed by it, as well as read by subsequent operations: the {\em item} is read by a subsequent dequeue and the {\em index} is read by subsequent enqueuers.  
To prevent reads of a location after it is flushed, an enqueuer in \ouq{} physically splits the node into two nodes. The first one is called \persistent{} and it is flushed and not accessed after the flush. It is only used during a recovery, for which its content is essential. It is allocated in the designated areas that the recovery will scan. The second node is denoted \volatile{} and it is not flushed and not used in a recovery. However, \volatile{} is accessed after the flush of \persistent{} and is utilized to expedite the normal operation on the object.
The {\em index} and {\em item} fields are placed in both \persistent{} and \volatile{}, with the two copies of each of them set to the same value. The enqueuer persists \persistent{}, while subsequent operations read the {\em index} and {\em item} from \volatile{}, thus adhering to our guideline. To enable access to the non-flushed fields, the queue's head and tail point to the \volatile{} part.

Each part of the node contains additional fields other than {\em index} and {\em item}:
The {\em linked} field is not accessed after the enqueuer performs the flush (except for during recovery), so there is no need to keep two copies of it, and it is placed in \persistent{} only.
The two following additional fields, which are not required in recovery, are placed in \volatile{}: {\em next}, and a pointer to the associated \persistent{} object, which the enqueuer sets for enabling the thread that reclaims the node later to locate \persistent{} and reclaim it together with \volatile{}. 

The recovery procedure of \ouq{} resurrects \persistent{} objects in the designated areas that are marked {\em linked} and have an {\em index} bigger than the head's index. It then allocates matching \volatile{} objects and links them in a linked list in the order induced by their indices.
This is implemented as follows. 

Let {\em headIndex} be the maximal index among the local head indices of all threads. These per-thread indices are not modified.
A dummy \persistent{} object is allocated and assigned the index {\em head\-Index}. Next, the recovery scans the designated areas and makes a list of recovered \persistent{} objects, which are those with a set {\em linked} flag and an {\em index} larger than {\em headIndex}. All other \persistent{} objects are reclaimed. 
Then, in order to construct a queue of \volatile{} objects, for each of the recovered \persistent{} objects, as well as for the dummy \persistent{}, the recovery allocates a \volatile{} object and sets a pointer from it to the associated \persistent{} object. In addition, the {\em index} and {\em item} of each \volatile{} are copied from the associated \persistent{}.
The \volatile{} objects are sorted by their indices, and their {\em next} pointers are set accordingly to create the queue. Finally, the queue's head and tail pointers are pointed at the first and last \volatile{} objects in the linked list.

\subsection{\olq{}}
\label{olq-overview}
% Rely on pred instead of next:
Transforming \liq{} to a queue with no access to flushed data is trickier and involves further modifications, since it is problematic to eliminate accesses to a node's {\em next} field after its flush. It is easier to avoid accessing a node's backward link {\em pred} after its flush, so we make the recovery rely on the node's {\em pred} instead of {\em next}. Accordingly, the recovery mechanism is reversed, so that instead of resurrecting a path of consecutive valid nodes reachable from the head (as \liq{} does), \olq{} resurrects a chain of consecutive nodes reachable from the tail by backward links, ending with the node succeeding the dummy node.
Similarly to \ouq{}, the queue node will be split into two nodes (\persistent{} and \volatile{}) so that the fields accessed after a flush (including the forward links) will not reside on the same cache line with the flushed fields (including the backward links).  

% Use index to identify stale nodes breaking the sequence:
Maintaining a single fence in each enqueue operation complicates the design of \olq{} further: An enqueuer needs to use a single fence to ensure the persistence of both all recently inserted nodes and the tail. Therefore, before the final fence, the tail might be already persisted while some nodes are not, which may cause the recovery to encounter stale nodes when walking from the tail backwards. The way we deal with this problem is to let the recovery identify stale nodes during the traversal. When a stale node is discovered, the recovery starts over from an older recorded value of the tail, and repeats this process until finding a recorded tail value from which the node succeeding the head is reachable through a chain of persisted nodes.
An {\em index} field placed in the nodes allows the recovery to identify stale nodes. These are nodes whose {\em index} value is nonconsecutive. This field is set in a new node by its enqueuer, to the index of the last node plus 1.

% recover HI:
The {\em index} field in nodes is also utilized to recover the head and the tail. As for the head, we cannot let dequeues flush the head pointer, because it will be accessed thereafter by following deq\-ueues. Like in \ouq{}, we assign a per-thread head index, which dequeues update with the head index and persist, and recover the head index as the maximum among these values in all threads. The recovery terminates its backward walk when it reach\-es a node with the head index plus 1.

% recover T:
We can also not let enqueues flush the tail, because it will be accessed thereafter by subsequent enqueues. To solve this, we assign a per-thread last-enqueue pointer (pointing to the last \persistent{} object enqueued by the thread), as well as a per-thread last-enqueue index. Note that a backward walk from a last-enqueue pointer of a thread that performed an enqueue during the crash, might pass through stale nodes, as the per-thread last-enqueue pointer and index might be persisted before some queue nodes are persisted. Thus, the recovery looks for the per-thread last-enqueue pointer pointing to the latest node {\em up to which all nodes are persisted}. The recovery starts the traversal from the node pointed to by the per-thread last-enqueue pointer with the maximum associated per-thread last-enqueue index among all threads, and if the {\em index} of this node is different from the associated last-enqueue index, or if nonconsecutive {\em index} values are encountered (each of these cases implies that the inspected node is stale), it restarts the walk from the next last-enqueue pointer candidate, which is the one with the next largest associated index, until it identifies a \persistent{} object from which it establishes a complete walk up to the node succeeding the head.

% including penultimate pointer and index:
The recovery scheme cannot be complete without dealing with the following rare scenario. All threads execute enqueues concurrently, the new last-enqueue pointer and index of them all are persisted in the memory, but then a crash occurs before any of the new nodes is persisted. In such a case, all last-enqueue pointers in all threads point to stale nodes, and the recovery will identify them as such. To restore a valid tail in this case, we assign {\em two} per-thread last-enqueue pointer and index, in which each thread keeps the details of both the last node enqueued by this thread and the penultimate node enqueued by this thread (up to which all queue nodes are definitely persisted by now because the penultimate enqueue was completed, including its fence instruction).  
The recovery sorts all last-enqueue indices (two of each thread) from largest to smallest and gathers their matching pointers to a single list of potential tail pointers. It attempts starting a backward walk from them, one after another. For each attempted tail pointer, if the index in the node it points to is different from the associated local enqueue index, or if a nonconsecutive index is encountered during the backward walk from it to the node with the recovered head index plus 1 (each of these cases implies that the index of the inspected node is stale) -- the recovery moves on to try the next potential tail.

% no need of initialized
An enqueuer sets the {\em index} of the new node after setting its {\em item} and {\em pred}, so based on  \Cref{release-fence-adequate-between-writes-to-same-cache-line}, when the recovery identifies the node's {\em index} as non-stale, it is guaranteed that its {\em item} and {\em pred} values are not stale.
In this new recovery scheme that uses {\em index} to detect stale nodes, an {\em initialized} field like in \liq{} is redundant. 

% Persistent, Volatile
Overall, the node's fields required in the recovery of \olq{} are {\em index}, {\em item} and {\em pred}. A node of \olq{} is composed of the following two parts: \persistent{} consists of the above mentioned fields, and \volatile{} consists copies of these fields for access after the flush of \persistent{}, as well as a {\em next} field that is not required in recovery, and a pointer to the associated \persistent{} object for its later reclamation.
Additional details of \olq{} appear in \Cref{olq-details}.

\subsection{Direct Write-Backs to Memory}\label{sec-opti}

The scheme described for \ouq{} replaces the global head index of \uq{}, which is read, written and persisted for an unbounded number of times, with local variables that are never read (except for during recovery). However, they are still written and persisted for an unbounded number of times: each dequeue operation writes and persists the local head index of its thread. A standard write to a value that is absent from the cache causes a fetch of the containing cache line from the memory. Thus, we wish to avoid such a write to a flushed (thus evicted) location.
Instead of a standard write, we issue a non-temporal write (using the \movnti{} instruction) of the local head index, which writes back the value to the memory without touching the cache. 
This way, \ouq{} optimally performs no access to flushed cache lines. 

To achieve this goal for \olq{} as well, we need to eliminate any access to its local variables.
The head index is handled just like in \ouq{}, using non-temporal writes. In addition, the local last-enqueue pointers and indices are also written and persisted for an unbounded number of times, and we update them too using non-temporal writes.

\section{Durable Linearizability}
\label{durable-linearizability-sketch}

To define linearization points for our queue algorithms, we first define some supporting terminology. We start with {\em \vlp{}s}, which match the standard linearization points of \msq{}, and are intuitively the steps applying the operations to the volatile queue. 
We also define a {\em \pp{}} for each operation, which marks the time from which the operation survives a crash. These two terms should basically be interpreted as: if an operation passes its \pp, then it is linearized at the time of its \vlp. If it does not reach its \pp, then it is not linearized in this execution.
Then, we derive the abstract state of the queue for each possible state of the queue's underlying list of nodes. 

\subsection{Linearization Points}
\label{linearization-points-subsection}

\begin{definition}[Volatile Linearization Point]
For each operation {\em op} in an execution {\em E} of the queue, we define its {\em \vlp{}} to be the same as {\em op}'s standard linearization point in \msq{}:
\begin{itemize}
    \item 
    Enqueue's \vlp{} is the \cas{} that links its new node (\volatile{} object in case of \ouq{} and \olq{}) to the last one.
    \item
    For a successful dequeue, its successful \cas{} that advances the queue's head is its \vlp{}.
    \item
    The \vlp{} of a failing dequeue is reading the {\em next} pointer of the dummy node (\volatile{} object in case of \ouq{} and \olq{}), which is later revealed to be \nul{}.
\end{itemize}
\end{definition}
An operation in {\em E} that does not reach its \vlp\ as defined above, does not have a \vlp{} (similarly to how not all operations in an execution have a linearization point).
Intuitively, an operation's \vlp{} is the step that applies the operation to the volatile queue.

\begin{definition}[Survival Point]\label{pp-def}
For each operation {\em op} in an execution {\em E} of the queue, we define a {\em \pp{}} as follows:
\begin{itemize}
    \item {\bf Successful Dequeue.}
    Let {\em op} be a successful dequeue that advances the head to point to {\em N} at moment $t$. 
    
    {\em op}'s \pp{} in \uq{} and \liq{} is the first (implicit or explicit) flush of the queue's head to the NVRAM after $t$, if a crash does not happen between $t$ and this flush (else, the dequeue operation does not have a \pp{}). (Note that the flushed value of the head could be pointing to {\em N} or a subsequent node in the queue).
    
    {\em op}'s \pp{} in \ouq{} and \olq{} is the first (implicit or explicit) flush of a per-thread head index to the NVRAM after $t$ with a value greater than or equal to {\em N}'s index, if a crash does not happen between $t$ and this flush (else, the dequeue operation does not have a \pp{}).
    
    \item {\bf Failing Dequeue.}
    Let {\em op} be a failing dequeue. Let {\em head} be the last value read off the queue's head, before discovering the queue is empty. This read is followed by {\em op}'s \vlp{}, where the {\em next} pointer in {\em head} is read and found \nul{}. Let $t_\ell$ be the time of this \vlp, and let us look back in time at the point $t$ where the value {\em head} was written to the queue's head. Let $t_p$ be the first time after $t$, where the content of the queue's head was flushed (implicitly or explicitly) to the memory, if a crash does not happen between $t$ and this flush (else, $t_p$ is undefined, and so is the \pp{} of the dequeue). Then {\em op}'s \pp{} in \uq{} and \liq{} is defined to be the later between $t_\ell$ and $t_p$. 
    
    {\em op}'s \pp{} in \ouq{} and \olq{} is defined similarly but using an alternative definition of $t_p$, as the moment of the first (implicit or explicit) flush of a per-thread head index to the NVRAM after $t$ with a value greater than or equal to {\em head}->{\em index}, if a crash does not happen between $t$ and this flush (else, $t_p$ is undefined, and so is the \pp{} of the dequeue).
    
    \item {\bf Enqueue.} 
    Let {\em op} be an enqueue operation that inserts {\em N} to the queue. By {\em N} we refer to a \texttt{Node} object linked to the queue in case of \uq{} and \liq{}, and to a \persistent{} object pointed to by a \volatile{} object that is linked to the queue in case of \ouq{} and \olq{}. Then the first of the following events to occur in {\em E} after the linking and before a crash occurs, is {\em op}'s \pp{} (if none of the following happens after the linking and before a crash, then the enqueue operation does not have a \pp{}):
    \begin{enumerate}
        \item\label{pp-enq-case-1} The queues differ in this event:
        \begin{itemize}
            \item For \uq{} and \ouq{}: An (implicit or explicit) flush of {\em N}'s {\em linked} field to the NVRAM after it is set to {\em true}.
            
            \item For \liq{}: 
            The first time when all of the following conditions have been met, for some node preceding {\em N}, denoted {\em dummy} (intuitively, {\em N} has become reachable from {\em dummy} and marked as initialized in the NVRAM view):
            \begin{enumerate}
                \item\label{pp-enq-case-1-liq-a} The queue's head has been flushed (implicitly or explicitly) to the NVRAM with a pointer to {\em dummy}. (Intuitively: {\em dummy} has become the queue's dummy node in the NVRAM view.)
                \item The underlying linked list of the queue connects {\em dummy} to {\em N}; and for each of the nodes along the way excluding {\em N}, its {\em next} field pointing to the subsequent node has been flushed (implicitly or explicitly) to the NVRAM. (Intuitively: {\em N} has been linked to the queue in the NV\-RAM view.)
                \item The setting of a {\em true} value to the {\em initialized} field in {\em N} reaches NVRAM by an (implicit or explicit) flush of {\em N}. (Intuitively: {\em N} has been marked as valid in the NVRAM.)
            \end{enumerate}
            
            \item For \olq{}: 
            The first time when all of the following conditions have been met, for some \persistent{} object preceding {\em N}, denoted {\em dummy}, and some \persistent{} object denoted {\em last} that is either {\em N} or a later \persistent{} object (intuitively, a backward path from the tail to the head through {\em N} became persistent):
            \begin{enumerate}
                \item Some per-thread head index has been flushed (implicitly or explicitly) to the NVRAM with the index of {\em dum\-my} (which means the head index will be recovered as {\em dummy}'s index or a bigger value).
                \item A last-enqueue pointer of some thread has been flushed (implicitly or explicitly) to the NVRAM with a pointer to {\em last}, and the associated last-enqueue index of that thread has been flushed to the NVRAM with the value {\em last.index}. (Intuitively: {\em last} has become a potential tail for the recovery.)
                \item The index of each \persistent{} object, from {\em last} backwards up to {\em dummy} excluding {\em dummy}, has been flush\-ed (implicitly or explicitly) to the NVRAM with its final value (namely, the indices of all these \persistent{} objects have been flushed with consecutive values).
            \end{enumerate}
        \end{itemize}

        \item\label{pp-enq-case-2} The \pp\ of a successful dequeue operation that dequeues the value inside {\em N}. 
    \end{enumerate}
\end{itemize}
\end{definition}
An operation in {\em E} that does not reach its defined-above \pp{} (in particular, a failing dequeue that does not reach both $t_\ell$ and $t_p$, and an enqueue that does not reach any of the two detailed points), either due to a crash or since the execution ends, does not have a \pp.

Intuitively, an operation's \pp{} is the flush that makes the operation survive a crash. The failing dequeue is somewhat different, as this operation does not modify the queue and we sometimes let its \pp\ be set to its \vlp, rather than a flush.  
Operations that reach a \pp{} are linearized even if a crash occurs after their \pp\ before they complete. 
Note that for our queues the \pp{} always happens when the \vlp{} has already occurred. 

\begin{definition}[Linearization Point]\label{def-linearization-point}
The {\em linearization point} of an operation {\em op} in an execution {\em E} of the queue, is defined to be its \vlp{} if {\em op} reaches a \pp{} in {\em E}. In this case, we say that {\em op} is {\em linearized}. Otherwise, {\em op} is not linearized, i.e., has no linearization point.
\end{definition}

\subsection{The Abstract State of the Queue}
\label{abstract-state-subsection}
We define the abstract state of the queue at each moment (including during the recovery) in a given execution of each queue. This state reflects the applying of all operations linearized so far in their linearization order.

\subsubsection{\uq{}}
The abstract head index in an execution of \uq{} is set to the value\footnote{To avoid confusion between the value in cache and the value in memory, we clarify that whenever a variable's value is mentioned, we refer to the last value written to the variable (regardless of whether it has reached the NVRAM).} of the {\em index} field in the queue's head except in an interval before a crash. Between the last flush of the head to the NVRAM before a crash and the crash, the value of the abstract head index is not modified. It remains the value that was flushed to the memory.  

The abstract state of the queue for execution {\em E} at moment $t$ is defined as all items in nodes with indices bigger than the current abstract head index, which were enqueued by linearized enqueues whose linearization points occurred prior to $t$, ordered by their enqueues' linearization order.

\subsubsection{\liq{}}
The abstract head of the queue in an execution {\em E} of \liq{} is defined similarly to the abstract head defined for \uq{}, but this time we look at the head pointer. 
We define the abstract head to be the queue's head value, except in an interval before a crash. From the last flush of the head (explicitly or implicitly) to the memory before a crash, until the crash, the abstract state of the head keeps the value flushed to the memory with no further abstract head state modifications in this interval. 

Consider an execution {\em E} of the queue and a moment $t$ during the execution, and consider the sequence of underlying list's nodes, starting with the dummy node pointed to by the abstract head, 
and ending with the first node along the chain whose {\em next} pointer is \nul{} or points to a node enqueued by a non linearized enqueue. Namely, we do not include nodes whose enqueues have not been linearized yet. 
The abstract state of the queue for {\em E} at $t$ is the sequence of items contained in all these nodes except for the first one (the dummy node).
Note that the abstract state of the queue is an empty sequence if and only if the {\em next} pointer of the dummy node is \nul{} or points to a node enqueued by a non linearized enqueue.

\subsubsection{\ouq{}}
The abstract head index of the queue in an execution {\em E} of \ouq{} is set to the value of the {\em index} field in the node pointed to by the queue's shared head, except in an interval enclosing a crash. 
Let {\em headIndex} be the biggest per-thread head index value flushed (explicitly or implicitly) to the NVRAM before the crash.
Between the moment a pointer to a node with the index {\em headIndex} is written to the queue's head and the moment the recovery procedure (that runs after the crash) terminates, the abstract head index keeps the value {\em headIndex}. 

The abstract state of the queue for execution {\em E} at moment $t$ is defined as all items in \persistent{} objects with indices bigger than the current abstract head index, which were enqueued by linearized enqueues whose linearization points occurred prior to $t$, ordered by their enqueues' linearization order.

\subsubsection{\olq{}}
The abstract head index of the queue in \olq{} is defined exactly like that of \ouq{}.
\olq{} is our only algorithm for which the abstract state of the queue depends also on the abstract state of the tail.
The abstract tail index in an execution {\em E} of \olq{} is set to the index of the node enqueued by the last linearized enqueue operation.
The abstract state of the queue is the sequence of items contained in the \persistent{} objects starting with the one enqueued by the last linearized enqueue and going through backward links until (including) the \persistent{} object with index bigger by 1 than the abstract head index, in reversed order; or an empty queue if the abstract tail index is not bigger than the abstract head index. 

\section{Lock-Freedom}
\label{Lock-Freedom-sketch}
Our queue algorithms are lock-free: An operation might fail to perform a \vlp{} only when another operation performs a conflicting \vlp{}, causing the original operation to retry in a new loop iteration. We argue that at a crash-free interval of execution, it is guaranteed that within a finite number of retries, some operation succeeds to reach not only a \vlp{}, but also a \pp{}, thus achieving a linearization point. Hence, system-wide progress is ensured.

The same basic argument applies to all operations of all presented queues:
A queue operation {\em op} branches backwards and starts a new loop iteration each time another operation performs an obstructing \vlp{}. If $op$ does not succeed to pursue a \vlp{} within {\em n} iterations, where {\em n} is the number of threads operating on the queue, then some other thread must have reached two \vlp{}s. This means it has completed the operation for which its first \vlp{} was reached, and persisted it before returning (to satisfy durable linearizability). Thus, this operation is linearized. Yet, its linearization point might have occurred before {\em op}'s execution, and we need to verify that some linearization point occurs during {\em op}'s execution.
We defer the details to \Cref{lock-freedom-section}.
\section{Memory Management}
\label{mm-section}
All queues evaluated in this paper (except for \ofq{} and \roq{} which were adopted from \cite{ramalhete2019onefile} and \cite{correia2020persistent} respectively as they are with their integrated memory manager), use the same version of epoch based reclamation for memory management, called {\em ssmem}. This memory manager is adopted from \cite{zuriel2019efficient}, which implements a durable extension of the mechanism presented by \cite{david2015asynchronized} for volatile memory. 
{\em ssmem} maintains designated areas in the heap memory for node allocation.
When a thread enqueues an item, it allocates a node from the next available space in these areas, or from a free list (to which dequeued nodes are inserted) if it is not empty.
The memory manager keeps a persistent list of all the areas it allocated throughout the execution. During recovery, free lists are reconstructed from the unused chunks in these areas.
Each thread in {\em ssmem} has its own allocator, operating on its separate designated areas and local free list, to avoid synchronization and reduce contention.
See \cite{zuriel2019efficient} for more details.

\section{Evaluation}
\label{performance-section}
\paragraph{Evaluated algorithms}
We compare to the durable queue in \cite{friedman2018persistent} as the most efficient lock-free durably linearizable queue algorithm known today. However, the queue as presented in \cite{friedman2018persistent} is built to satisfy more than just durable linearizability. It contains a mechanism for retrieving previously obtained results after a crash, which is not required by durable linearizability, and is not provided by other durable data structures \cite{david2018log,zuriel2019efficient}. 
To put all these data structures on the same level of guarantees, we remove the additional mechanism from~\cite{friedman2018persistent}, obtaining a thinner version of the original durable queue that executes faster, a version we denote \dmsq{}. 
Comparison to the exact original queue from~\cite{friedman2018persistent} would yield better performance for us, but would not be fair. 
The extra mechanism in \cite{friedman2018persistent} can be easily added to the versions we propose (with the corresponding additional cost). 

In addition, we compare to a persistent queue implementation resulting from applying the general construction of Izraelevitz~\cite{izraelevitz2016linearizability} to \msq{}. We also compare to the persistent queue version obtained by NVTraverse \cite{friedman2020nvtraverse}, which resembles \iq{} since the traversal phase in \msq{} is empty, hence, the operations access directly the critical point, being the head or tail. The only difference between the two versions is that \nvtq{} does not issue a fence after a flush that follows a read or \cas{} instruction.
To complement the comparison, we compare to queues produced by wrapping a sequential queue implementation with a persistent transactional memory (PTM):
\ofq{}, produced using the OneFile lock-free PTM \cite{ramalhete2019onefile}, and \roq{}, produced using the RedoOpt PTM \cite{correia2020persistent}.

\begin{figure}
  \centering 
  \includegraphics[width=.45\textwidth]{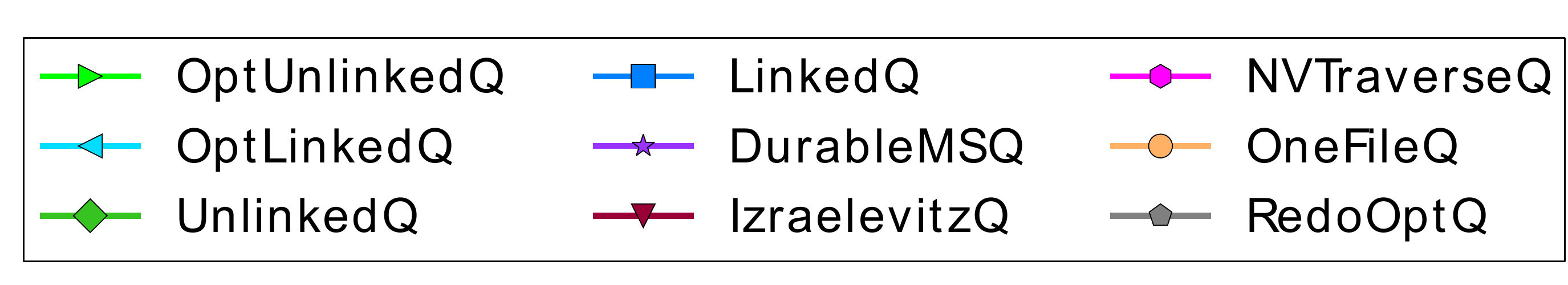}
  \includegraphics[width=.5\textwidth]{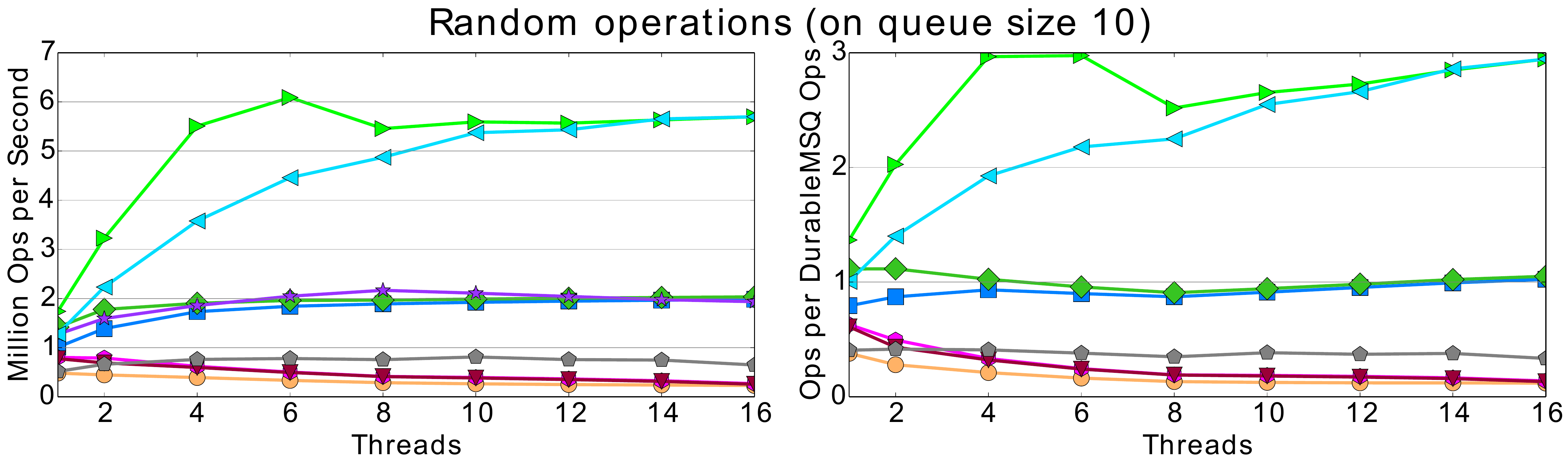}
  \includegraphics[width=.5\textwidth]{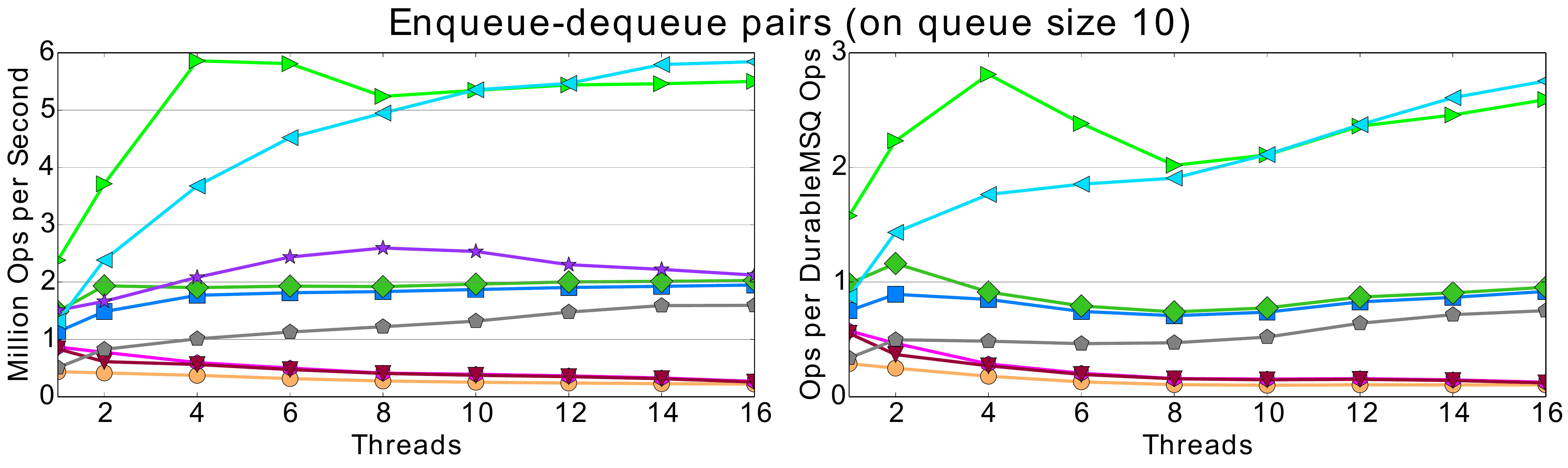}
  \includegraphics[width=.5\textwidth]{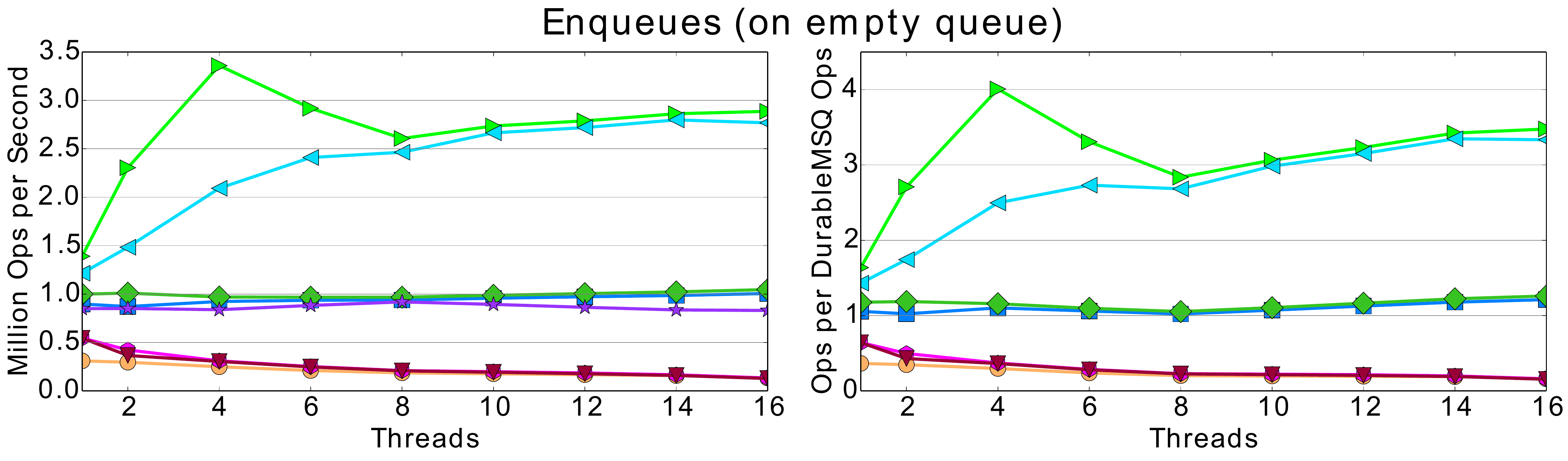}
  \includegraphics[width=.5\textwidth]{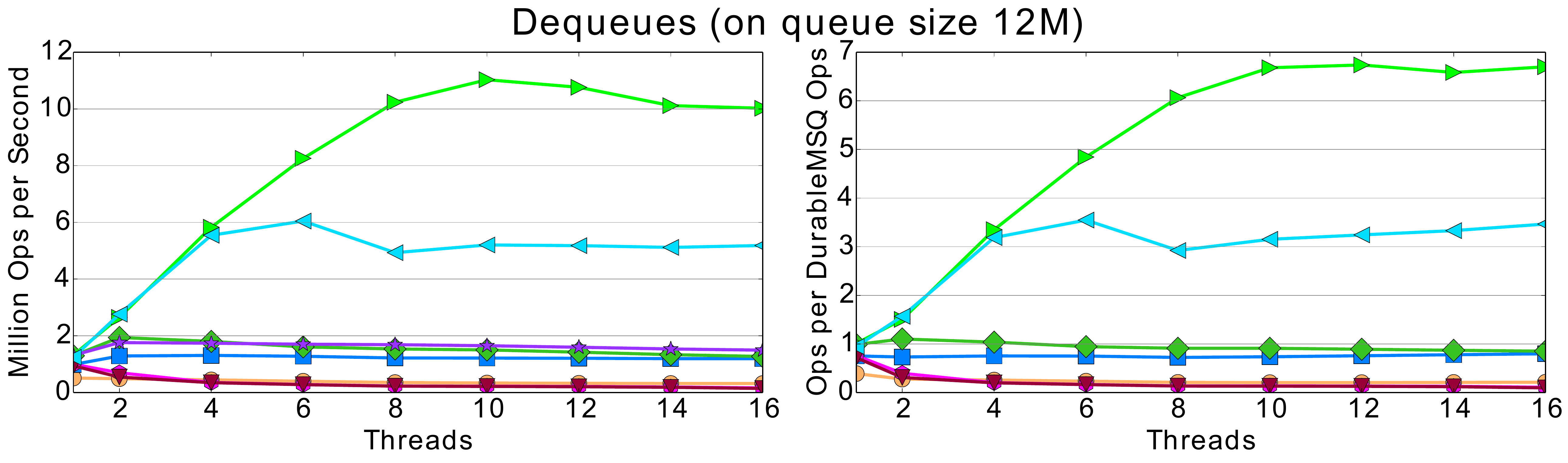}
  \includegraphics[width=.5\textwidth]{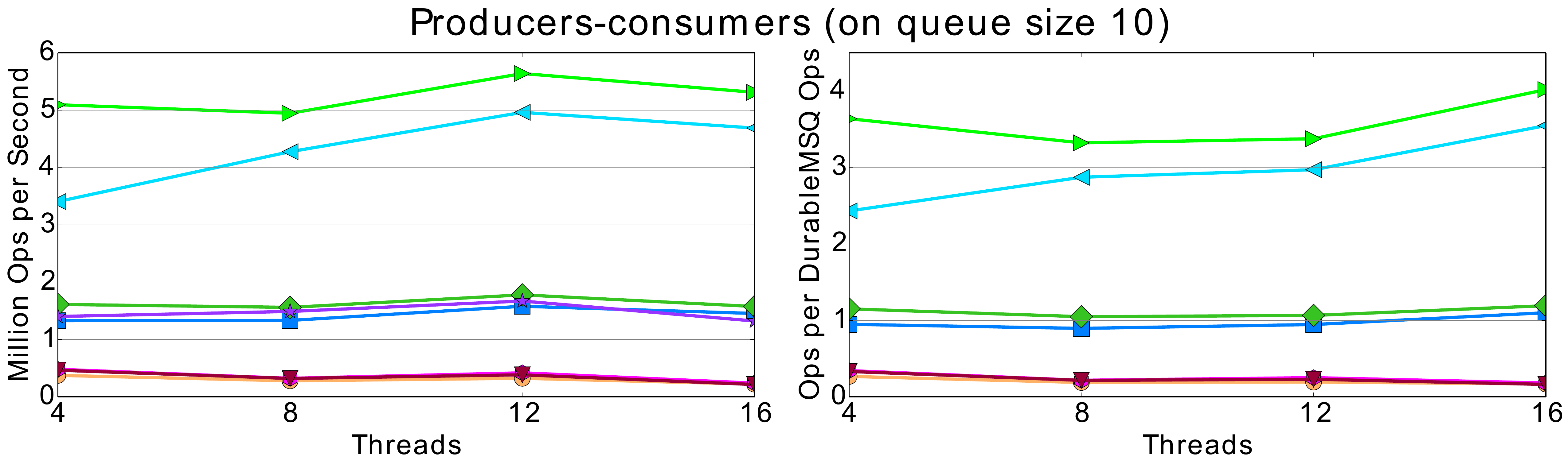}
  \caption{Measurement results}
  \label{fig:results}
\end{figure}

\paragraph{Platform}
The queues were implemented in C++ and compiled using the g++ (GCC) compiler version 9.3.0 with a -O3 optimization level. 
We conducted our experiments on a machine running Linux (Ubuntu 18.04) equipped with 2 Intel Xeon Gold 6234 3.3GHz processors with 8 cores each.
In experiments with up to 8 threads, each thread was attached to a different core of the same processor.
In experiments with more than 8 threads in which the ninth and on threads were attached to the second processor, NUMA effects kick in impeding scalability and reducing performance, but the trends remain the same (\ouq{} performs best, \olq{} is second best).
To avoid NUMA effects, we utilize hyper-threading (SMT) on a single processor for measurements of more than 8 threads reported in \Cref{fig:results}: we attach the $(8+i)^{th}$ thread to the second virtual core on the same physical core as the $i^{th}$ thread.

The machine has an L1 data cache of 32KB and an L2 cache of 1MB per core, and an L3 cache of 25MB per processor. It has 1.5TB of NVRAM (Intel Optane DC Persistent Memory), organized as 128GB DIMMs (6 per processor).
The machine uses the NVRAM in App-Direct Mode Interleaved in our configuration.
\clwb{} is utilized as a flush instruction, \sfence{} as a store fence and \movnti{} as a write-back to memory (non-temporal store) instruction.

\paragraph{Methodology}
In each experiment, the queue is initialized with a certain number of enqueued items, and then operations are applied to it, for five seconds unless specified otherwise.  
Each data point $[x,y]$ in the graphs represents the average result of 10 experiments. In each experiment, $x$ threads performed operations concurrently. 
The left graphs depict the throughput, namely, number of operations applied to each queue per second by the threads altogether. The right graphs depict the throughput ratio between each queue and the baseline \dmsq{}.

We ran various workloads following prior works (see \Cref{fig:results}): 
In the first workload, operations were randomly chosen to be enqueue or dequeue (50-50 uniform distribution) following \cite{yang2016wait,hoffman2007baskets,ladan2004optimistic}.
In the second workload, each thread ran enqueue-dequeue pairs, following measurements in \cite{michael1996simple,friedman2018persistent,ramalhete2019onefile,yang2016wait,morrison2013fast,fatourou2012revisiting,hoffman2007baskets,ladan2004optimistic}.
Next, we ran producers only (performing enqueues) on an empty queue. We also ran consumers only (performing dequeues) on a queue of size 12M following \cite{ostrovsky2020scaling} for 1 second.
At last, we ran a mixed producer-consumer workload, loosely following \cite{ostrovsky2020scaling,haas2013distributed,kirsch2013fast}. Here, unlike in other workloads, the threads did not run for a preset amount of time, but rather executed a preset number of operations: one quarter of the threads performed 1M dequeues and then 1M enqueues, and the rest performed 1M enqueues and then 1M dequeues. This is intended to ensure that the queue is not drained, as enqueues are slower than dequeues.
The initial queue in the presented graphs in the first, second and last workloads is of size 10. An initial size of 10K yields similar results (as we do not traverse the entire queue, but only touch the front and rear of the queue).
RedoOpt is evaluated only in the first two workloads since we had problems running it on the other workloads.

\paragraph{Results}
Our first two queue designs, \uq{} and \liq{}, perform better than \dmsq{} for some workloads and worse for others. They do not gain an advantage over \dmsq{} although performing minimum fences, due to accesses to flushed cache lines.
Our efficient transformations that avoid such accesses, \ouq{} and \olq{}, outperform all other queues including \dmsq{}, the state-of-the-art durable queue, in nearly all experiments.
For example, \ouq{} runs more than twice faster than \dmsq{} for nearly all workloads with more than one thread.
\iq{} is substantially slower than \dmsq{} and our queues, as expected from a universal construction that places many more fences than the tailor-made queues. \nvtq{}, which is similar to \iq{}, shows nearly identical performance.
The transactional approach of \ofq{} and \roq{} results in reduced performance as transactions impose additional overhead over a short operation.

\section{Conclusion}
\label{conclusion-section}
In this paper we presented a new guideline for designing efficient durable algorithms suitable for the current architecture: reducing accesses to flushed memory. We demonstrated the advantage of following this guideline with durable queues. We first present novel queues that abide only to the known guideline of minimizing the fence count, meeting the theoretical lower bound on the number of fences from \cite{cohen2018inherent}, executing only one blocking fence per operation. 
\uq{} does not persist the links, but rather allocates the nodes on designated areas and adds an ordering mechanism, so the recovery procedure can look for valid nodes of the queue in the designated areas and order them correctly.
\liq{} uses a validity scheme on the queue nodes to inform the recovery algorithm which nodes are adequate for recovery, and adds a backward link to the queue’s underlying structure to allow enqueues to persist previous enqueues efficiently. 
These queues do not beat state-of-the-art queues in spite of issuing fewer fences. 
We then amended these queues to achieve zero accesses to flushed memory while still maintaining a single blocking fence per operation.
The resulted queues demonstrate a significant performance improvement on the Intel Optane NVRAM over state-of-the-art durable queues, showing that, at least in our context, the second amendment is desirable.

\bibliographystyle{ACM-Reference-Format}
\balance 
\bibliography{refs}


\begin{thebibliography}{57}

%%% the Web address.

\ifx \showCODEN    \undefined \def \showCODEN     #1{\unskip}     \fi
\ifx \showDOI      \undefined \def \showDOI       #1{#1}\fi
\ifx \showISBNx    \undefined \def \showISBNx     #1{\unskip}     \fi
\ifx \showISBNxiii \undefined \def \showISBNxiii  #1{\unskip}     \fi
\ifx \showISSN     \undefined \def \showISSN      #1{\unskip}     \fi
\ifx \showLCCN     \undefined \def \showLCCN      #1{\unskip}     \fi
\ifx \shownote     \undefined \def \shownote      #1{#1}          \fi
\ifx \showarticletitle \undefined \def \showarticletitle #1{#1}   \fi
\ifx \showURL      \undefined \def \showURL       {\relax}        \fi
% invisible to TeX
\providecommand\bibfield[2]{#2}
\providecommand\bibinfo[2]{#2}
\providecommand\natexlab[1]{#1}
\providecommand\showeprint[2][]{arXiv:#2}

\bibitem[\protect\citeauthoryear{Aguilera and Fr{\o}lund}{Aguilera and
  Fr{\o}lund}{2003}]%
        {aguilera2003strict}
\bibfield{author}{\bibinfo{person}{Marcos~K Aguilera} {and}
  \bibinfo{person}{Svend Fr{\o}lund}.} \bibinfo{year}{2003}\natexlab{}.
\newblock \showarticletitle{Strict linearizability and the power of aborting}.
\newblock \bibinfo{journal}{\emph{Technical Report HPL-2003-241}}
  (\bibinfo{year}{2003}).
\newblock
\urldef\tempurl%
\url{https://hpl.hp.com/techreports/2003/HPL-2003-241.html}
\showURL{%
\tempurl}


\bibitem[\protect\citeauthoryear{Akinaga and Shima}{Akinaga and Shima}{2010}]%
        {akinaga2010resistive}
\bibfield{author}{\bibinfo{person}{Hiroyuki Akinaga} {and}
  \bibinfo{person}{Hisashi Shima}.} \bibinfo{year}{2010}\natexlab{}.
\newblock \showarticletitle{Resistive random access memory ({ReRAM}) based on
  metal oxides}.
\newblock \bibinfo{journal}{\emph{Proc. IEEE}} \bibinfo{volume}{98},
  \bibinfo{number}{12} (\bibinfo{year}{2010}).
\newblock
\urldef\tempurl%
\url{https://doi.org/10.1109/JPROC.2010.2070830}
\showDOI{\tempurl}


\bibitem[\protect\citeauthoryear{Berryhill, Golab, and Tripunitara}{Berryhill
  et~al\mbox{.}}{2015}]%
        {berryhill2015robust}
\bibfield{author}{\bibinfo{person}{Ryan Berryhill}, \bibinfo{person}{Wojciech
  Golab}, {and} \bibinfo{person}{Mahesh Tripunitara}.}
  \bibinfo{year}{2015}\natexlab{}.
\newblock \showarticletitle{Robust shared objects for non-volatile main
  memory}. In \bibinfo{booktitle}{\emph{OPODIS}}.
\newblock
\urldef\tempurl%
\url{https://doi.org/10.4230/LIPIcs.OPODIS.2015.20}
\showDOI{\tempurl}


\bibitem[\protect\citeauthoryear{Brais and Rudoff}{Brais and Rudoff}{2021}]%
        {stackoverflow-confirming-atomic-cache-line-writeback}
\bibfield{author}{\bibinfo{person}{Hadi Brais} {and} \bibinfo{person}{Andy
  Rudoff}.} \bibinfo{year}{2021}\natexlab{}.
\newblock \bibinfo{booktitle}{\emph{Reply to On x86-64, is the "movnti" or
  "movntdq" instruction atomic when system crash?}}
\newblock
\urldef\tempurl%
\url{https://stackoverflow.com/a/65587308/7289606}
\showURL{%
\tempurl}


\bibitem[\protect\citeauthoryear{Bu, Dong, Yi, Zang, and Chen}{Bu
  et~al\mbox{.}}{2021}]%
        {bu2021revisiting}
\bibfield{author}{\bibinfo{person}{Heng Bu}, \bibinfo{person}{Ming-Kai Dong},
  \bibinfo{person}{Ji-Fei Yi}, \bibinfo{person}{Bin-Yu Zang}, {and}
  \bibinfo{person}{Hai-Bo Chen}.} \bibinfo{year}{2021}\natexlab{}.
\newblock \showarticletitle{Revisiting persistent indexing structures on Intel
  Optane DC persistent memory}.
\newblock \bibinfo{journal}{\emph{Journal of Computer Science and Technology}}
  \bibinfo{volume}{36}, \bibinfo{number}{1} (\bibinfo{year}{2021}).
\newblock
\urldef\tempurl%
\url{https://doi.org/10.1007/s11390-020-9871-0}
\showDOI{\tempurl}


\bibitem[\protect\citeauthoryear{Chakrabarti, Boehm, and Bhandari}{Chakrabarti
  et~al\mbox{.}}{2014}]%
        {chakrabarti2014atlas}
\bibfield{author}{\bibinfo{person}{Dhruva~R Chakrabarti},
  \bibinfo{person}{Hans-J Boehm}, {and} \bibinfo{person}{Kumud Bhandari}.}
  \bibinfo{year}{2014}\natexlab{}.
\newblock \showarticletitle{Atlas: Leveraging locks for non-volatile memory
  consistency}.
\newblock \bibinfo{journal}{\emph{ACM SIGPLAN Notices}} \bibinfo{volume}{49},
  \bibinfo{number}{10} (\bibinfo{year}{2014}).
\newblock
\urldef\tempurl%
\url{https://doi.org/10.1145/2714064.2660224}
\showDOI{\tempurl}


\bibitem[\protect\citeauthoryear{Coburn, Caulfield, Akel, Grupp, Gupta, Jhala,
  and Swanson}{Coburn et~al\mbox{.}}{2011}]%
        {coburn2011nv}
\bibfield{author}{\bibinfo{person}{Joel Coburn}, \bibinfo{person}{Adrian~M
  Caulfield}, \bibinfo{person}{Ameen Akel}, \bibinfo{person}{Laura~M Grupp},
  \bibinfo{person}{Rajesh~K Gupta}, \bibinfo{person}{Ranjit Jhala}, {and}
  \bibinfo{person}{Steven Swanson}.} \bibinfo{year}{2011}\natexlab{}.
\newblock \showarticletitle{{NV-Heaps}: Making persistent objects fast and safe
  with next-generation, non-volatile memories}. In
  \bibinfo{booktitle}{\emph{ASPLOS}}.
\newblock
\urldef\tempurl%
\url{https://doi.org/10.1145/1961295.1950380}
\showDOI{\tempurl}


\bibitem[\protect\citeauthoryear{Cohen, Friedman, and Larus}{Cohen
  et~al\mbox{.}}{2017}]%
        {cohen2017efficient}
\bibfield{author}{\bibinfo{person}{Nachshon Cohen}, \bibinfo{person}{Michal
  Friedman}, {and} \bibinfo{person}{James~R Larus}.}
  \bibinfo{year}{2017}\natexlab{}.
\newblock \showarticletitle{Efficient logging in non-volatile memory by
  exploiting coherency protocols}.
\newblock \bibinfo{journal}{\emph{PACMPL}} \bibinfo{volume}{1},
  \bibinfo{number}{OOPSLA} (\bibinfo{year}{2017}).
\newblock
\urldef\tempurl%
\url{https://doi.org/10.1145/3133891}
\showDOI{\tempurl}


\bibitem[\protect\citeauthoryear{Cohen, Guerraoui, and Zablotchi}{Cohen
  et~al\mbox{.}}{2018}]%
        {cohen2018inherent}
\bibfield{author}{\bibinfo{person}{Nachshon Cohen}, \bibinfo{person}{Rachid
  Guerraoui}, {and} \bibinfo{person}{Igor Zablotchi}.}
  \bibinfo{year}{2018}\natexlab{}.
\newblock \showarticletitle{The inherent cost of remembering consistently}. In
  \bibinfo{booktitle}{\emph{SPAA}}.
\newblock
\urldef\tempurl%
\url{https://doi.org/10.1145/3210377.3210400}
\showDOI{\tempurl}


\bibitem[\protect\citeauthoryear{Correia, Felber, and Ramalhete}{Correia
  et~al\mbox{.}}{2018}]%
        {correia2018romulus}
\bibfield{author}{\bibinfo{person}{Andreia Correia}, \bibinfo{person}{Pascal
  Felber}, {and} \bibinfo{person}{Pedro Ramalhete}.}
  \bibinfo{year}{2018}\natexlab{}.
\newblock \showarticletitle{Romulus: Efficient algorithms for persistent
  transactional memory}. In \bibinfo{booktitle}{\emph{SPAA}}.
\newblock
\urldef\tempurl%
\url{https://doi.org/10.1145/3210377.3210392}
\showDOI{\tempurl}


\bibitem[\protect\citeauthoryear{Correia, Felber, and Ramalhete}{Correia
  et~al\mbox{.}}{2020}]%
        {correia2020persistent}
\bibfield{author}{\bibinfo{person}{Andreia Correia}, \bibinfo{person}{Pascal
  Felber}, {and} \bibinfo{person}{Pedro Ramalhete}.}
  \bibinfo{year}{2020}\natexlab{}.
\newblock \showarticletitle{Persistent memory and the rise of universal
  constructions}. In \bibinfo{booktitle}{\emph{EuroSys}}.
\newblock
\urldef\tempurl%
\url{https://doi.org/10.1145/3342195.3387515}
\showDOI{\tempurl}


\bibitem[\protect\citeauthoryear{David, Dragojevic, Guerraoui, and
  Zablotchi}{David et~al\mbox{.}}{2018}]%
        {david2018log}
\bibfield{author}{\bibinfo{person}{Tudor David}, \bibinfo{person}{Aleksandar
  Dragojevic}, \bibinfo{person}{Rachid Guerraoui}, {and} \bibinfo{person}{Igor
  Zablotchi}.} \bibinfo{year}{2018}\natexlab{}.
\newblock \showarticletitle{Log-free concurrent data structures}. In
  \bibinfo{booktitle}{\emph{USENIX ATC}}.
\newblock
\urldef\tempurl%
\url{https://usenix.org/conference/atc18/presentation/david}
\showURL{%
\tempurl}


\bibitem[\protect\citeauthoryear{David, Guerraoui, and Trigonakis}{David
  et~al\mbox{.}}{2015}]%
        {david2015asynchronized}
\bibfield{author}{\bibinfo{person}{Tudor David}, \bibinfo{person}{Rachid
  Guerraoui}, {and} \bibinfo{person}{Vasileios Trigonakis}.}
  \bibinfo{year}{2015}\natexlab{}.
\newblock \showarticletitle{Asynchronized concurrency: The secret to scaling
  concurrent search data structures}. In \bibinfo{booktitle}{\emph{ASPLOS}}.
\newblock
\urldef\tempurl%
\url{https://doi.org/10.1145/2786763.2694359}
\showDOI{\tempurl}


\bibitem[\protect\citeauthoryear{Fatourou and Kallimanis}{Fatourou and
  Kallimanis}{2012}]%
        {fatourou2012revisiting}
\bibfield{author}{\bibinfo{person}{Panagiota Fatourou} {and}
  \bibinfo{person}{Nikolaos~D Kallimanis}.} \bibinfo{year}{2012}\natexlab{}.
\newblock \showarticletitle{Revisiting the combining synchronization
  technique}. In \bibinfo{booktitle}{\emph{PPoPP}}.
\newblock
\urldef\tempurl%
\url{https://doi.org/10.1145/2370036.2145849}
\showDOI{\tempurl}


\bibitem[\protect\citeauthoryear{Friedman, Ben-David, Wei, Blelloch, and
  Petrank}{Friedman et~al\mbox{.}}{2020}]%
        {friedman2020nvtraverse}
\bibfield{author}{\bibinfo{person}{Michal Friedman}, \bibinfo{person}{Naama
  Ben-David}, \bibinfo{person}{Yuanhao Wei}, \bibinfo{person}{Guy~E Blelloch},
  {and} \bibinfo{person}{Erez Petrank}.} \bibinfo{year}{2020}\natexlab{}.
\newblock \showarticletitle{{NVTraverse}: In {NVRAM} data structures, the
  destination is more important than the journey}. In
  \bibinfo{booktitle}{\emph{PLDI}}.
\newblock
\urldef\tempurl%
\url{https://doi.org/10.1145/3385412.3386031}
\showDOI{\tempurl}


\bibitem[\protect\citeauthoryear{Friedman, Herlihy, Marathe, and
  Petrank}{Friedman et~al\mbox{.}}{2018}]%
        {friedman2018persistent}
\bibfield{author}{\bibinfo{person}{Michal Friedman}, \bibinfo{person}{Maurice
  Herlihy}, \bibinfo{person}{Virendra Marathe}, {and} \bibinfo{person}{Erez
  Petrank}.} \bibinfo{year}{2018}\natexlab{}.
\newblock \showarticletitle{A persistent lock-free queue for non-volatile
  memory}. In \bibinfo{booktitle}{\emph{PPoPP}}.
\newblock
\urldef\tempurl%
\url{https://doi.org/10.1145/3178487.3178490}
\showDOI{\tempurl}


\bibitem[\protect\citeauthoryear{Friedman, Petrank, and Ramalhete}{Friedman
  et~al\mbox{.}}{2021}]%
        {friedman2021mirror}
\bibfield{author}{\bibinfo{person}{Michal Friedman}, \bibinfo{person}{Erez
  Petrank}, {and} \bibinfo{person}{Pedro Ramalhete}.}
  \bibinfo{year}{2021}\natexlab{}.
\newblock \showarticletitle{Mirror: Making lock-free data structures
  persistent}. In \bibinfo{booktitle}{\emph{PLDI}}.
\newblock
\urldef\tempurl%
\url{https://doi.org/10.1145/3453483.3454105}
\showDOI{\tempurl}


\bibitem[\protect\citeauthoryear{Guerraoui and Levy}{Guerraoui and
  Levy}{2004}]%
        {guerraoui2004robust}
\bibfield{author}{\bibinfo{person}{Rachid Guerraoui} {and}
  \bibinfo{person}{Ron~R Levy}.} \bibinfo{year}{2004}\natexlab{}.
\newblock \showarticletitle{Robust emulations of shared memory in a
  crash-recovery model}. In \bibinfo{booktitle}{\emph{ICDCS}}.
\newblock
\urldef\tempurl%
\url{https://doi.org/10.1109/ICDCS.2004.1281605}
\showDOI{\tempurl}


\bibitem[\protect\citeauthoryear{Haas, Lippautz, Henzinger, Payer, Sokolova,
  Kirsch, and Sezgin}{Haas et~al\mbox{.}}{2013}]%
        {haas2013distributed}
\bibfield{author}{\bibinfo{person}{Andreas Haas}, \bibinfo{person}{Michael
  Lippautz}, \bibinfo{person}{Thomas~A Henzinger}, \bibinfo{person}{Hannes
  Payer}, \bibinfo{person}{Ana Sokolova}, \bibinfo{person}{Christoph~M Kirsch},
  {and} \bibinfo{person}{Ali Sezgin}.} \bibinfo{year}{2013}\natexlab{}.
\newblock \showarticletitle{Distributed queues in shared memory: Multicore
  performance and scalability through quantitative relaxation}. In
  \bibinfo{booktitle}{\emph{CF}}.
\newblock
\urldef\tempurl%
\url{https://doi.org/10.1145/2482767.2482789}
\showDOI{\tempurl}


\bibitem[\protect\citeauthoryear{Hao}{Hao}{2019}]%
        {clwb-blog-post}
\bibfield{author}{\bibinfo{person}{Xiangpeng Hao}.}
  \bibinfo{year}{2019}\natexlab{}.
\newblock \bibinfo{booktitle}{\emph{Is CLWB actually implemented?}}
\newblock
\urldef\tempurl%
\url{https://blog.haoxp.xyz/posts/is-clwb-implemented}
\showURL{%
\tempurl}


\bibitem[\protect\citeauthoryear{Herlihy}{Herlihy}{1991}]%
        {herlihy1991wait}
\bibfield{author}{\bibinfo{person}{Maurice Herlihy}.}
  \bibinfo{year}{1991}\natexlab{}.
\newblock \showarticletitle{Wait-free synchronization}.
\newblock \bibinfo{journal}{\emph{TOPLAS}} \bibinfo{volume}{13},
  \bibinfo{number}{1} (\bibinfo{year}{1991}).
\newblock
\urldef\tempurl%
\url{https://doi.org/10.1145/114005.102808}
\showDOI{\tempurl}


\bibitem[\protect\citeauthoryear{Herlihy and Wing}{Herlihy and Wing}{1990}]%
        {herlihy1990linearizability}
\bibfield{author}{\bibinfo{person}{Maurice Herlihy} {and}
  \bibinfo{person}{Jeannette~M. Wing}.} \bibinfo{year}{1990}\natexlab{}.
\newblock \showarticletitle{Linearizability: A correctness condition for
  concurrent objects}.
\newblock \bibinfo{journal}{\emph{TOPLAS}} \bibinfo{volume}{12},
  \bibinfo{number}{3} (\bibinfo{year}{1990}).
\newblock
\urldef\tempurl%
\url{https://doi.org/10.1145/78969.78972}
\showDOI{\tempurl}


\bibitem[\protect\citeauthoryear{Hoffman, Shalev, and Shavit}{Hoffman
  et~al\mbox{.}}{2007}]%
        {hoffman2007baskets}
\bibfield{author}{\bibinfo{person}{Moshe Hoffman}, \bibinfo{person}{Ori
  Shalev}, {and} \bibinfo{person}{Nir Shavit}.}
  \bibinfo{year}{2007}\natexlab{}.
\newblock \showarticletitle{The baskets queue}. In
  \bibinfo{booktitle}{\emph{OPODIS}}.
\newblock
\urldef\tempurl%
\url{https://doi.org/10.1007/978-3-540-77096-1_29}
\showDOI{\tempurl}


\bibitem[\protect\citeauthoryear{IBM}{IBM}{[n.d.]}]%
        {ibmmq}
\bibfield{author}{\bibinfo{person}{IBM}.} \bibinfo{year}{[n.d.]}\natexlab{}.
\newblock \bibinfo{booktitle}{\emph{{IBM MQ}}}.
\newblock
\urldef\tempurl%
\url{https://ibm.com/software/products/en/ibm-mq}
\showURL{%
\tempurl}


\bibitem[\protect\citeauthoryear{Intel}{Intel}{2019}]%
        {3DX}
\bibfield{author}{\bibinfo{person}{Intel}.} \bibinfo{year}{2019}\natexlab{}.
\newblock \bibinfo{booktitle}{\emph{{3D XPoint}\texttrademark: A breakthrough
  in non-volatile memory technology}}.
\newblock
\urldef\tempurl%
\url{https://intel.com/content/www/us/en/architecture-and-technology/intel-micron-3d-xpoint-webcast.html}
\showURL{%
\tempurl}


\bibitem[\protect\citeauthoryear{Intel}{Intel}{2020}]%
        {IntelSDM}
\bibfield{author}{\bibinfo{person}{Intel}.} \bibinfo{year}{2020}\natexlab{}.
\newblock \bibinfo{booktitle}{\emph{Intel\textregistered 64 and {IA}-32
  architectures software developer's manual}}.
\newblock
\urldef\tempurl%
\url{https://software.intel.com/content/dam/develop/external/us/en/documents-tps/325462-sdm-vol-1-2abcd-3abcd.pdf}
\showURL{%
\tempurl}


\bibitem[\protect\citeauthoryear{Izraelevitz, Mendes, and Scott}{Izraelevitz
  et~al\mbox{.}}{2016}]%
        {izraelevitz2016linearizability}
\bibfield{author}{\bibinfo{person}{Joseph Izraelevitz},
  \bibinfo{person}{Hammurabi Mendes}, {and} \bibinfo{person}{Michael~L Scott}.}
  \bibinfo{year}{2016}\natexlab{}.
\newblock \showarticletitle{Linearizability of persistent memory objects under
  a full-system-crash failure model}. In \bibinfo{booktitle}{\emph{DISC}}.
\newblock
\urldef\tempurl%
\url{https://doi.org/10.1007/978-3-662-53426-7_23}
\showDOI{\tempurl}


\bibitem[\protect\citeauthoryear{Kalia, Andersen, and Kaminsky}{Kalia
  et~al\mbox{.}}{2020}]%
        {kalia2020challenges}
\bibfield{author}{\bibinfo{person}{Anuj Kalia}, \bibinfo{person}{David
  Andersen}, {and} \bibinfo{person}{Michael Kaminsky}.}
  \bibinfo{year}{2020}\natexlab{}.
\newblock \showarticletitle{Challenges and solutions for fast remote persistent
  memory access}. In \bibinfo{booktitle}{\emph{SoCC}}.
\newblock
\urldef\tempurl%
\url{https://doi.org/10.1145/3419111.3421294}
\showDOI{\tempurl}


\bibitem[\protect\citeauthoryear{Kirsch, Lippautz, and Payer}{Kirsch
  et~al\mbox{.}}{2013}]%
        {kirsch2013fast}
\bibfield{author}{\bibinfo{person}{Christoph~M Kirsch},
  \bibinfo{person}{Michael Lippautz}, {and} \bibinfo{person}{Hannes Payer}.}
  \bibinfo{year}{2013}\natexlab{}.
\newblock \showarticletitle{Fast and scalable, lock-free k-{FIFO} queues}. In
  \bibinfo{booktitle}{\emph{PaCT}}.
\newblock
\urldef\tempurl%
\url{https://doi.org/10.1007/978-3-642-39958-9_18}
\showDOI{\tempurl}


\bibitem[\protect\citeauthoryear{Kolli, Pelley, Saidi, Chen, and Wenisch}{Kolli
  et~al\mbox{.}}{2016}]%
        {kolli2016high}
\bibfield{author}{\bibinfo{person}{Aasheesh Kolli}, \bibinfo{person}{Steven
  Pelley}, \bibinfo{person}{Ali Saidi}, \bibinfo{person}{Peter~M Chen}, {and}
  \bibinfo{person}{Thomas~F Wenisch}.} \bibinfo{year}{2016}\natexlab{}.
\newblock \showarticletitle{High-performance transactions for persistent
  memories}. In \bibinfo{booktitle}{\emph{ASPLOS}}.
\newblock
\urldef\tempurl%
\url{https://doi.org/10.1145/2872362.2872381}
\showDOI{\tempurl}


\bibitem[\protect\citeauthoryear{Ladan-Mozes and Shavit}{Ladan-Mozes and
  Shavit}{2004}]%
        {ladan2004optimistic}
\bibfield{author}{\bibinfo{person}{Edya Ladan-Mozes} {and} \bibinfo{person}{Nir
  Shavit}.} \bibinfo{year}{2004}\natexlab{}.
\newblock \showarticletitle{An optimistic approach to lock-free {FIFO} queues}.
  In \bibinfo{booktitle}{\emph{Distributed Computing}}.
\newblock
\urldef\tempurl%
\url{https://doi.org/10.1007/s00446-007-0050-0}
\showDOI{\tempurl}


\bibitem[\protect\citeauthoryear{Lea}{Lea}{2009}]%
        {lea2009java}
\bibfield{author}{\bibinfo{person}{Doug Lea}.} \bibinfo{year}{2009}\natexlab{}.
\newblock \bibinfo{title}{The Java concurrency package ({JSR}-166)}.
\newblock
\newblock


\bibitem[\protect\citeauthoryear{Marathe, Mishra, Trivedi, Huang, Zaghloul,
  Kashyap, Seltzer, Harris, Byan, Bridge, et~al\mbox{.}}{Marathe
  et~al\mbox{.}}{2018}]%
        {marathe2018persistent}
\bibfield{author}{\bibinfo{person}{Virendra Marathe}, \bibinfo{person}{Achin
  Mishra}, \bibinfo{person}{Amee Trivedi}, \bibinfo{person}{Yihe Huang},
  \bibinfo{person}{Faisal Zaghloul}, \bibinfo{person}{Sanidhya Kashyap},
  \bibinfo{person}{Margo Seltzer}, \bibinfo{person}{Tim Harris},
  \bibinfo{person}{Steve Byan}, \bibinfo{person}{Bill Bridge}, {et~al\mbox{.}}}
  \bibinfo{year}{2018}\natexlab{}.
\newblock \showarticletitle{Persistent memory transactions}.
\newblock \bibinfo{journal}{\emph{arXiv preprint}} (\bibinfo{year}{2018}).
\newblock
\showeprint{1804.00701}


\bibitem[\protect\citeauthoryear{Memaripour, Izraelevitz, and
  Swanson}{Memaripour et~al\mbox{.}}{2020}]%
        {memaripour2020pronto}
\bibfield{author}{\bibinfo{person}{Amirsaman Memaripour},
  \bibinfo{person}{Joseph Izraelevitz}, {and} \bibinfo{person}{Steven
  Swanson}.} \bibinfo{year}{2020}\natexlab{}.
\newblock \showarticletitle{Pronto: Easy and fast persistence for volatile data
  structures}. In \bibinfo{booktitle}{\emph{ASPLOS}}.
\newblock
\urldef\tempurl%
\url{https://doi.org/10.1145/3373376.3378456}
\showDOI{\tempurl}


\bibitem[\protect\citeauthoryear{Michael and Scott}{Michael and Scott}{1996}]%
        {michael1996simple}
\bibfield{author}{\bibinfo{person}{Maged~M. Michael} {and}
  \bibinfo{person}{Michael~L. Scott}.} \bibinfo{year}{1996}\natexlab{}.
\newblock \showarticletitle{Simple, fast, and practical non-blocking and
  blocking concurrent queue algorithms}. In \bibinfo{booktitle}{\emph{PODC}}.
\newblock
\urldef\tempurl%
\url{https://doi.org/10.1145/248052.248106}
\showDOI{\tempurl}


\bibitem[\protect\citeauthoryear{Morrison and Afek}{Morrison and Afek}{2013}]%
        {morrison2013fast}
\bibfield{author}{\bibinfo{person}{Adam Morrison} {and} \bibinfo{person}{Yehuda
  Afek}.} \bibinfo{year}{2013}\natexlab{}.
\newblock \showarticletitle{Fast concurrent queues for x86 processors}. In
  \bibinfo{booktitle}{\emph{PPoPP}}.
\newblock
\urldef\tempurl%
\url{https://doi.org/10.1145/2442516.2442527}
\showDOI{\tempurl}


\bibitem[\protect\citeauthoryear{Oracle}{Oracle}{[n.d.]}]%
        {tuxedomq}
\bibfield{author}{\bibinfo{person}{Oracle}.} \bibinfo{year}{[n.d.]}\natexlab{}.
\newblock \bibinfo{booktitle}{\emph{{Oracle Tuxedo Message Queue}}}.
\newblock
\urldef\tempurl%
\url{https://docs.oracle.com/cd/E35855_01/otmq/docs12c/overview/overview.html}
\showURL{%
\tempurl}


\bibitem[\protect\citeauthoryear{Ostrovsky and Morrison}{Ostrovsky and
  Morrison}{2020}]%
        {ostrovsky2020scaling}
\bibfield{author}{\bibinfo{person}{Or Ostrovsky} {and} \bibinfo{person}{Adam
  Morrison}.} \bibinfo{year}{2020}\natexlab{}.
\newblock \showarticletitle{Scaling concurrent queues by using {HTM} to profit
  from failed atomic operations}. In \bibinfo{booktitle}{\emph{PPoPP}}.
\newblock
\urldef\tempurl%
\url{https://doi.org/10.1145/3332466.3374511}
\showDOI{\tempurl}


\bibitem[\protect\citeauthoryear{Raad, Wickerson, Neiger, and Vafeiadis}{Raad
  et~al\mbox{.}}{2020}]%
        {raad2020persistency}
\bibfield{author}{\bibinfo{person}{Azalea Raad}, \bibinfo{person}{John
  Wickerson}, \bibinfo{person}{Gil Neiger}, {and} \bibinfo{person}{Viktor
  Vafeiadis}.} \bibinfo{year}{2020}\natexlab{}.
\newblock \showarticletitle{Persistency semantics of the {I}ntel-x86
  architecture}.
\newblock \bibinfo{journal}{\emph{PACMPL}} \bibinfo{volume}{4},
  \bibinfo{number}{POPL} (\bibinfo{year}{2020}).
\newblock
\urldef\tempurl%
\url{https://doi.org/10.1145/3371079}
\showDOI{\tempurl}


\bibitem[\protect\citeauthoryear{Ramalhete, Correia, Felber, and
  Cohen}{Ramalhete et~al\mbox{.}}{2019}]%
        {ramalhete2019onefile}
\bibfield{author}{\bibinfo{person}{Pedro Ramalhete}, \bibinfo{person}{Andreia
  Correia}, \bibinfo{person}{Pascal Felber}, {and} \bibinfo{person}{Nachshon
  Cohen}.} \bibinfo{year}{2019}\natexlab{}.
\newblock \showarticletitle{{OneFile}: A wait-free persistent transactional
  memory}. In \bibinfo{booktitle}{\emph{DSN}}.
\newblock
\urldef\tempurl%
\url{https://doi.org/10.1109/DSN.2019.00028}
\showDOI{\tempurl}


\bibitem[\protect\citeauthoryear{Raoux, Burr, Breitwisch, Rettner, Chen,
  Shelby, Salinga, Krebs, Chen, Lung, et~al\mbox{.}}{Raoux
  et~al\mbox{.}}{2008}]%
        {raoux2008phase}
\bibfield{author}{\bibinfo{person}{Simone Raoux}, \bibinfo{person}{Geoffrey~W
  Burr}, \bibinfo{person}{Matthew~J Breitwisch}, \bibinfo{person}{Charles~T
  Rettner}, \bibinfo{person}{Y-C Chen}, \bibinfo{person}{Robert~M Shelby},
  \bibinfo{person}{Martin Salinga}, \bibinfo{person}{Daniel Krebs},
  \bibinfo{person}{S-H Chen}, \bibinfo{person}{H-L Lung}, {et~al\mbox{.}}}
  \bibinfo{year}{2008}\natexlab{}.
\newblock \showarticletitle{Phase-change random access memory: A scalable
  technology}.
\newblock \bibinfo{journal}{\emph{IBM Journal of Research and Development}}
  \bibinfo{volume}{52}, \bibinfo{number}{4.5} (\bibinfo{year}{2008}).
\newblock
\urldef\tempurl%
\url{https://doi.org/10.1147/rd.524.0465}
\showDOI{\tempurl}


\bibitem[\protect\citeauthoryear{Rudoff}{Rudoff}{2019}]%
        {google-group-1-confirming-atomic-cache-line-writeback}
\bibfield{author}{\bibinfo{person}{Andy Rudoff}.}
  \bibinfo{year}{2019}\natexlab{}.
\newblock \bibinfo{booktitle}{\emph{Reply to How to use CLWB instructions}}.
\newblock
\urldef\tempurl%
\url{https://groups.google.com/g/pmem/c/R8H3sKq9sLQ/m/ltL7Kng4BAAJ}
\showURL{%
\tempurl}


\bibitem[\protect\citeauthoryear{Rudoff}{Rudoff}{2020}]%
        {google-group-2-confirming-atomic-cache-line-writeback}
\bibfield{author}{\bibinfo{person}{Andy Rudoff}.}
  \bibinfo{year}{2020}\natexlab{}.
\newblock \bibinfo{booktitle}{\emph{Reply to 8 byte atomicity \& larger store
  operations}}.
\newblock
\urldef\tempurl%
\url{https://groups.google.com/g/pmem/c/6\_5daOuEI00/m/nY\_mtKd0CAAJ}
\showURL{%
\tempurl}


\bibitem[\protect\citeauthoryear{Scargall}{Scargall}{2020}]%
        {scargall2020programming}
\bibfield{author}{\bibinfo{person}{Steve Scargall}.}
  \bibinfo{year}{2020}\natexlab{}.
\newblock \bibinfo{booktitle}{\emph{Programming persistent memory: A
  comprehensive guide for developers}}.
\newblock \bibinfo{publisher}{Springer Nature}.
\newblock
\urldef\tempurl%
\url{https://doi.org/10.1007/978-1-4842-4932-1}
\showDOI{\tempurl}


\bibitem[\protect\citeauthoryear{Sela, Herlihy, and Petrank}{Sela
  et~al\mbox{.}}{2021}]%
        {sela2021linearizability}
\bibfield{author}{\bibinfo{person}{Gal Sela}, \bibinfo{person}{Maurice
  Herlihy}, {and} \bibinfo{person}{Erez Petrank}.}
  \bibinfo{year}{2021}\natexlab{}.
\newblock \showarticletitle{Brief announcement: Linearizability: A typo}. In
  \bibinfo{booktitle}{\emph{PODC}}.
\newblock
\urldef\tempurl%
\url{https://doi.org/10.1145/3465084.3467944}
\showDOI{\tempurl}


\bibitem[\protect\citeauthoryear{Sela and Petrank}{Sela and Petrank}{2021}]%
        {sela2021durablequeues}
\bibfield{author}{\bibinfo{person}{Gal Sela} {and} \bibinfo{person}{Erez
  Petrank}.} \bibinfo{year}{2021}\natexlab{}.
\newblock \showarticletitle{Durable queues: The second amendment}. In
  \bibinfo{booktitle}{\emph{SPAA}}.
\newblock
\urldef\tempurl%
\url{https://doi.org/10.1145/3409964.3461791}
\showDOI{\tempurl}


\bibitem[\protect\citeauthoryear{SNIA}{SNIA}{2017}]%
        {SNIA}
\bibfield{author}{\bibinfo{person}{SNIA}.} \bibinfo{year}{2017}\natexlab{}.
\newblock \bibinfo{booktitle}{\emph{{NVM} Programming Model ({NPM})}}.
\newblock
\urldef\tempurl%
\url{https://snia.org/sites/default/files/technical_work/final/NVMProgrammingModel_v1.2.pdf}
\showURL{%
\tempurl}


\bibitem[\protect\citeauthoryear{Software}{Software}{[n.d.]}]%
        {rabbitmq}
\bibfield{author}{\bibinfo{person}{Pivotal Software}.}
  \bibinfo{year}{[n.d.]}\natexlab{}.
\newblock \bibinfo{booktitle}{\emph{{RabbitMQ}}}.
\newblock
\urldef\tempurl%
\url{https://rabbitmq.com}
\showURL{%
\tempurl}


\bibitem[\protect\citeauthoryear{team}{team}{[n.d.]}]%
        {pmem}
\bibfield{author}{\bibinfo{person}{Intel~PMDK team}.}
  \bibinfo{year}{[n.d.]}\natexlab{}.
\newblock \bibinfo{booktitle}{\emph{Persistent memory programming}}.
\newblock
\urldef\tempurl%
\url{https://pmem.io}
\showURL{%
\tempurl}


\bibitem[\protect\citeauthoryear{van Renen, Vogel, Leis, Neumann, and
  Kemper}{van Renen et~al\mbox{.}}{2019}]%
        {van2019persistent}
\bibfield{author}{\bibinfo{person}{Alexander van Renen}, \bibinfo{person}{Lukas
  Vogel}, \bibinfo{person}{Viktor Leis}, \bibinfo{person}{Thomas Neumann},
  {and} \bibinfo{person}{Alfons Kemper}.} \bibinfo{year}{2019}\natexlab{}.
\newblock \showarticletitle{Persistent memory {I/O} primitives}. In
  \bibinfo{booktitle}{\emph{DaMoN}}.
\newblock
\urldef\tempurl%
\url{https://doi.org/10.1145/3329785.3329930}
\showDOI{\tempurl}


\bibitem[\protect\citeauthoryear{Volos, Tack, and Swift}{Volos
  et~al\mbox{.}}{2011}]%
        {volos2011mnemosyne}
\bibfield{author}{\bibinfo{person}{Haris Volos}, \bibinfo{person}{Andres~Jaan
  Tack}, {and} \bibinfo{person}{Michael~M Swift}.}
  \bibinfo{year}{2011}\natexlab{}.
\newblock \showarticletitle{Mnemosyne: Lightweight persistent memory}. In
  \bibinfo{booktitle}{\emph{ASPLOS}}.
\newblock
\urldef\tempurl%
\url{https://doi.org/10.1145/1950365.1950379}
\showDOI{\tempurl}


\bibitem[\protect\citeauthoryear{Wen, Cai, Du, Jenkins, Valpey, and Scott}{Wen
  et~al\mbox{.}}{2020}]%
        {wen2020montage}
\bibfield{author}{\bibinfo{person}{Haosen Wen}, \bibinfo{person}{Wentao Cai},
  \bibinfo{person}{Mingzhe Du}, \bibinfo{person}{Louis Jenkins},
  \bibinfo{person}{Benjamin Valpey}, {and} \bibinfo{person}{Michael~L Scott}.}
  \bibinfo{year}{2020}\natexlab{}.
\newblock \showarticletitle{Montage: A general system for buffered durably
  linearizable data structures}.
\newblock \bibinfo{journal}{\emph{arXiv preprint}} (\bibinfo{year}{2020}).
\newblock
\showeprint{2009.13701}


\bibitem[\protect\citeauthoryear{Wu, Ren, and Li}{Wu et~al\mbox{.}}{2019}]%
        {wu2019architecture}
\bibfield{author}{\bibinfo{person}{Kai Wu}, \bibinfo{person}{Jie Ren}, {and}
  \bibinfo{person}{Dong Li}.} \bibinfo{year}{2019}\natexlab{}.
\newblock \showarticletitle{Architecture-aware, high performance transaction
  for persistent memory}.
\newblock \bibinfo{journal}{\emph{arXiv preprint}} (\bibinfo{year}{2019}).
\newblock
\showeprint{1903.06226}


\bibitem[\protect\citeauthoryear{Yang and Mellor-Crummey}{Yang and
  Mellor-Crummey}{2016}]%
        {yang2016wait}
\bibfield{author}{\bibinfo{person}{Chaoran Yang} {and} \bibinfo{person}{John
  Mellor-Crummey}.} \bibinfo{year}{2016}\natexlab{}.
\newblock \showarticletitle{A wait-free queue as fast as fetch-and-add}. In
  \bibinfo{booktitle}{\emph{PPoPP}}.
\newblock
\urldef\tempurl%
\url{https://doi.org/10.1145/2851141.2851168}
\showDOI{\tempurl}


\bibitem[\protect\citeauthoryear{Yang, Kim, Hoseinzadeh, Izraelevitz, and
  Swanson}{Yang et~al\mbox{.}}{2020}]%
        {yang2020empirical}
\bibfield{author}{\bibinfo{person}{Jian Yang}, \bibinfo{person}{Juno Kim},
  \bibinfo{person}{Morteza Hoseinzadeh}, \bibinfo{person}{Joseph Izraelevitz},
  {and} \bibinfo{person}{Steve Swanson}.} \bibinfo{year}{2020}\natexlab{}.
\newblock \showarticletitle{An empirical guide to the behavior and use of
  scalable persistent memory}. In \bibinfo{booktitle}{\emph{FAST}}.
\newblock


\bibitem[\protect\citeauthoryear{Zardoshti, Zhou, Liu, and Spear}{Zardoshti
  et~al\mbox{.}}{2019}]%
        {zardoshti2019optimizing}
\bibfield{author}{\bibinfo{person}{Pantea Zardoshti}, \bibinfo{person}{Tingzhe
  Zhou}, \bibinfo{person}{Yujie Liu}, {and} \bibinfo{person}{Michael Spear}.}
  \bibinfo{year}{2019}\natexlab{}.
\newblock \showarticletitle{Optimizing persistent memory transactions}. In
  \bibinfo{booktitle}{\emph{PACT}}.
\newblock
\urldef\tempurl%
\url{https://doi.org/10.1109/PACT.2019.00025}
\showDOI{\tempurl}


\bibitem[\protect\citeauthoryear{Zuriel, Friedman, Sheffi, Cohen, and
  Petrank}{Zuriel et~al\mbox{.}}{2019}]%
        {zuriel2019efficient}
\bibfield{author}{\bibinfo{person}{Yoav Zuriel}, \bibinfo{person}{Michal
  Friedman}, \bibinfo{person}{Gali Sheffi}, \bibinfo{person}{Nachshon Cohen},
  {and} \bibinfo{person}{Erez Petrank}.} \bibinfo{year}{2019}\natexlab{}.
\newblock \showarticletitle{Efficient lock-free durable sets}.
\newblock \bibinfo{journal}{\emph{PACMPL}} \bibinfo{volume}{3},
  \bibinfo{number}{OOPSLA} (\bibinfo{year}{2019}).
\newblock
\urldef\tempurl%
\url{https://doi.org/10.1145/3360554}
\showDOI{\tempurl}


\end{thebibliography}

\clearpage

\appendix
\section{\liq{} Details}\label{liq-details}
The \liq{} algorithm appears in \Cref{fig:liq-impl}. Next, we describe its operations in detail.
\begin{figure}
\caption{\liq{} implementation}\label{fig:liq-impl}
\vspace{0.5\baselineskip}
\begin{minipage}{.49\textwidth}
\begin{lstlisting}
class Node
    Item* item
    atomic<Node*> next
    atomic<Node*> pred
    bool initialized
\end{lstlisting}
\begin{lstlisting}
@\underline{Item* Dequeue()}@
    while (true)
        head = Head@\label{liq-deq-read-head}@
        headNext = head->next@\label{liq-deq-read-headNext}@
        if (headNext == NULL)
            FLUSH(&Head); SFENCE@\label{liq-failing-deq-persist-head}@
            return NULL@\label{liq-failing-deq-return-null}@
        if (CAS(&Head, head, headNext)@\label{liq-advance-head}@
            dequeuedItem = headNext->item
            if (nodeToPersistAndRetire[tid])
                FLUSH(&nodeToPersistAndRetire[tid]->initialized)@\label{liq-flush-invalid-successful-deq}@
            FLUSH(&Head)@\label{liq-successful-deq-persist-head-flush}@
            SFENCE@\label{liq-successful-deq-persist-head-fence}@
            headNext->pred = NULL@\label{liq-deq-set-pred-null}@
            if (nodeToPersistAndRetire[tid])
                retire(nodeToPersistAndRetire[tid])@\label{liq-retire-successful-deq}@
            head->initialized = false@\label{liq-set-uninitialized}@
            nodeToPersistAndRetire[tid] = head@\label{liq-save-node-to-reclaim}@
            return dequeuedItem
\end{lstlisting}
\end{minipage}\hfill
\begin{minipage}{.49\textwidth}
\begin{lstlisting}
@\underline{FlushNotPersistedSuffix(notPersisted)}@
    do@\label{liq-flushNotPersistedSuffix-begin}@
        FLUSH(notPersisted)
        notPersisted = notPersisted->pred
    while (notPersisted != NULL);@\label{liq-flushNotPersistedSuffix-end}@
@\underline{Enqueue(item)}@
    newNode = allocNode()@\label{liq-alloc-node}@
    newNode->item = item
    newNode->next = NULL@\label{liq-enq-set-next-null}@
    newNode->initialized = true@\label{liq-set-initialized}@
    while (true)
        tail = Tail
        if (tail->next == NULL)
            newNode->pred = tail@\label{liq-set-pred}@
            if (CAS(&tail->next, NULL, newNode))@\label{liq-link}@
                FlushNotPersistedSuffix(newNode)@\label{liq-flushNotPersistedSuffix}@
                SFENCE@\label{liq-fence}@
                CAS(&Tail, tail, newNode)@\label{liq-advance-tail}@
                // All nodes preceding newNode are persistent
                newNode->pred = NULL@\label{liq-enq-set-pred-null}@
                break
        CAS(&Tail, tail, tail->next)@\label{liq-enq-assisting-cas-tail}@
\end{lstlisting}
\end{minipage}
\end{figure}

\subsection{The Enqueue Operation}\label{liq-enq-description}
The enqueue operation first allocates a node denoted {\em newNode} from the memory manager and initializes its data (\Crefrange{liq-alloc-node}{liq-enq-set-next-null}).
Then it sets {\em newNode}'s {\em initialized} flag (\Cref{liq-set-initialized}).
In what follows, the enqueue operation attempts to link {\em newNode} to the last node (\Cref{liq-link}).
Note that it might have just linked a node whose data is not persisted in NVRAM. If the link to {\em newNode} is written back to NVRAM (which could happen implicitly due to a cache eviction) and then a crash occurs, the recovery would have to deal with reaching a node with stale data. The correctness is maintained using the {\em initialized} flag and a matching recovery procedure:
The {\em initialized} flag is used as a stamp indicating to the recovery that {\em newNode} is initialized. 
Relying on \Cref{release-fence-adequate-between-writes-to-same-cache-line}, the order of writing {\em initialized} after {\em newNode}'s data is preserved in NVRAM.
Accordingly, the recovery resurrects only nodes with the {\em initialized} flag set. This guarantees that it resurrects only nodes with persisted data.
If a crash occurs after the link to {\em newNode} is flushed to NVRAM and before {\em newNode}'s data is written back to NVRAM, then {\em newNode}'s {\em initialized} flag must be unset in NVRAM, thus, the recovery will ignore it (and all nodes linked after it).

The recovery procedure resurrects all nodes reachable from the head through a path of consecutive nodes with the {\em initialized} flag set.
Since durable linearizability allows the recovery procedure to ignore enqueue operations that are concurrent with a crash and elide their nodes from the queue, the fact that the recovery procedure ignores nodes of ongoing enqueues that were linked after a node with an unset {\em initialized}, does not break durable linearizability.
However, after an enqueue operation completes, the recovery procedure must not discard it, even if earlier nodes belong to incomplete enqueue operations.
To this end, after successfully linking {\em newNode}, the enqueue operation ensures that the path of nodes leading from the head to {\em newNode} is persisted (\Crefrange{liq-flushNotPersistedSuffix}{liq-fence}).
This could be achieved naively by flushing all nodes from the head until {\em newNode}. To save redundant flushes, an enqueuer avoids flushing a prefix of queue's nodes that are guaranteed to be already flushed. Instead, it flushes only a suffix of queue's nodes that are not guaranteed to be persistent.
To identify the relevant suffix, we place backward links in the nodes, but we remove a backward link when we know that all previous nodes in the queue have already been persisted.
The backward links preserve the following invariant: all queue nodes (starting from the current queue's head) that precede a node with a nullified backward link have all relevant content (their item, set {\em initialized} flag and a non-\nul{} forward link) persisted. 

To maintain a backward path connecting the linked list's nodes that should be flushed, an enqueuer links a node with a backward link pointing to the previous node (\Cref{liq-set-pred}).
After linking {\em newNode}, its enqueuer traverses the queue from {\em newNode} backwards using the backward links, until reaching a \nul{} backward link, and flushes the content of all traversed nodes (including {\em newNode} itself) (\Crefrange{liq-flushNotPersistedSuffix-begin}{liq-flushNotPersistedSuffix-end}).
Finally, it issues a single \sfence{} to block until all these flushes complete (\Cref{liq-fence}).
By the above-mentioned invariant, all nodes starting from the current head and preceding this suffix of nodes, are persistent. 
Now, as this suffix is persisted as well, the data of all nodes preceding {\em newNode} starting from the current head is guaranteed to be persistent. As an optimization to prevent future enqueues from flushing these nodes, the enqueuer then sets {\em newNode}'s backward link to \nul{} (\Cref{liq-enq-set-pred-null}). Thus, each enqueue operation that reaches {\em newNode} from now on, during its backward walk, would not need to traverse the preceding persistent nodes. Note that backward links are not used in the recovery and there is no need to explicitly flush them.

To complete the enqueue operation, the tail is advanced to point to {\em newNode} (\Cref{liq-advance-tail}).
Like in the original \msq{}, a concurrent enqueue might prevent the enqueue's linking. In this case, the enqueuer tries to assist and advance the tail to point to the node enqueued by the obstructing enqueue (\Cref{liq-enq-assisting-cas-tail}), before starting a new attempt to enqueue its own item.

We note that, as an optimization on x86 platforms, the \sfence{} in \Cref{liq-fence} can be eliminated, because the following \cas{} instruction serves as an \sfence{} guaranteeing completion of previous flushes. In the measured implementations of all algorithms, each \sfence{} preceding a \cas{} is eliminated.
We did not include this optimization in the paper's pseudocode for clarity.

\subsection{The Dequeue Operation}\label{liq-deq-description}
The dequeue operation attempts to extract the oldest item, placed in the node subsequent to the dummy node.
If the queue is empty when the dequeue operation takes effect, it returns \nul{}.
But before returning, the failing dequeue must persist the head (\Cref{liq-failing-deq-persist-head}), to ensure that previous ongoing dequeues that emptied the queue are persistent. 
Otherwise, if a crash occurs after the failing dequeue returns, the previous dequeues might be discarded. This would break durable linearizability, since it will be impossible to linearize the completed failing dequeue correctly as applied to an empty queue, without the previous dequeues being linearized beforehand. 

If the queue is not empty, the dequeuer attempts to advance the head by one node (\Cref{liq-advance-head}), and on success -- returns the oldest item to the caller. On failure it retries the whole scheme.
Before returning, the dequeuer persists the head (\Crefrange{liq-successful-deq-persist-head-flush}{liq-successful-deq-persist-head-fence}), to comply with durable linearizability, which requires that completed operations be linearized.  

Each dequeue makes sure that the dummy node from which it advances the head will be unreachable by future operations, so that the next successful dequeue by the same thread will safely return  this dummy node to the memory manager.  
To make it unreachable by backward walks (of enqueue operations that will try to identify a not persisted suffix), the dequeuer disconnects the backward link from the new dummy head to the previous one (\Cref{liq-deq-set-pred-null}).
In addition, persisting the head guarantees that the previous dummy node will be unreachable by future operations even in case of a crash.

A successful dequeue does not simply return the previous dummy node (i.e., the node from which the previous successful dequeue by the same thread has advanced the head) to the memory manager as it is.
Recall from \Cref{liq-overview-subsection} that we must make sure that newly allocated nodes have their {\em initialized} flags reset. 
The {\em initialized} flag placed in each node is used by its enqueuer to signal to the recovery when the node's data is persisted. Suppose a node is erroneously allocated in an enqueue operation with a set {\em initialized} flag.
After the enqueue operation links the node, the link to the node might be implicitly flushed to the NVRAM, and -- before the node's data is persisted -- a crash might follow. The recovery procedure would then find the linked node, containing stale data including a set {\em initialized} flag, and would erroneously interpret the node with the stale content as part of the queue.
To prevent this scenario, enqueuers could unset the {\em initialized} flag after the node's allocation and then persist it before initializing its data, but this incurs an additional fence.
Instead, we make sure that a node is always allocated with an {\em initialized} flag persistently unset.
Next we explain how we ensure that.

If we allocate nodes from the operating system, we would get nodes with arbitrary content, possibly with the {\em initialized} field set. Instead, we implement a memory manager that maintains large designated areas from which all node allocations are performed. 

First, we explain how nodes, allocated from the designated areas for the first time, are allocated with a persistently unset {\em initialized} value.
If the number of nodes required by the program is known in advance, then on program startup, the memory manager may allocate a sufficiently large designated area for nodes from the operating system, zero its content to make all nodes marked as not initialized, and then persist it in NVRAM (by placing asynchronous flushes of the whole area accompanied by a single \sfence{}). This guarantees that when the memory manager allocates a node for the first time, its {\em initialized} field is unset.
If the number of required nodes is unknown in advance, each time a designated area is depleted, the memory manager may allocate a new area from the operating system, and initialize it in a similar manner using a single \sfence{}.

It remains to explain how nodes, reallocated from the designated areas after reclamation, are allocated with an {\em initialized} flag persistently unset.
The dequeue operation and the recovery procedure return nodes to the memory manager, hence, they are responsible to return them with an {\em initialized} flag persistently unset.

Starting with dequeue, a successful dequeuer could unset the {\em initialized} flag of the dummy node from which it has advanced the head and then perform additional flush and \sfence{} to persist the unset {\em initialized} flag before returning the node to the memory manager.
However, to achieve the fence lower bound of a single \sfence{} per operation, \liq{} takes a different approach. 

The persistence of the previous dummy node's {\em initialized} flag is accomplished through piggybacking on the next successful dequeue's \sfence{}, which this thread is anyhow going to execute (in \Cref{liq-successful-deq-persist-head-fence}). More precisely, the dequeuer sets the previous dummy node's {\em initialized} flag to {\em false} (\Cref{liq-set-uninitialized}) after the queue's head persistently points to a subsequent node. 
The dequeuer thread postpones the reclamation of this previous dummy node, and keeps the node locally in a {\em nodeToPersistAndRetire} array (\Cref{liq-save-node-to-reclaim}). This array consists of a pointer cell per thread, each cell lying in another cache line to avoid false sharing. Each thread may access its cell using its thread ID as an index. 
In the next successful dequeue execution of the same thread, right before its \sfence{}, the {\em initialized} flag of the node we kept aside is flushed (\Cref{liq-flush-invalid-successful-deq}). 
After the fence completes, the node may be returned to the memory manager (\Cref{liq-retire-successful-deq}). 

As for the recovery, as detailed in \Cref{liq-recovery-subsection}, for each node with a set {\em initialized} flag that it returns to the memory manager -- the recovery unsets the flag and flushes it. A single \sfence{} placed in the end of the recovery ensures that these flags are unset in the memory.

\subsection{Recovery}
\label{liq-recovery-subsection}
The recovery procedure of \liq{}, running after a crash, resurrects all nodes reachable from the head through a path of consecutive nodes with the {\em initialized} flag set. It does so by leaving the queue's head as it is and reconstructing the queue as follows.
\begin{enumerate}
    \item\label{liq-recovery-head-init-unset} If the {\em initialized} flag of the dummy node (namely, the node pointed to by the head) is unset, the recovery procedure sets the dummy node's {\em next} to \nul{} and then sets its {\em initialized} flag. The order of the last two writes ensures (based on \Cref{release-fence-adequate-between-writes-to-same-cache-line}) a proper recovery from a possible crash in the midst of the current recovery. The tail is set to point to the dummy node as well. 
    \item\label{liq-recovery-head-init-set} Otherwise (the dummy node's {\em initialized} is set) --
    \begin{enumerate}
        \item\label{liq-recovery-last-node-with-next-null} The recovery procedure traverses the nodes starting with the one pointed to by the head, until it reaches either a node whose {\em next} value is \nul{}, or a node with an unset {\em initialized}. In the first case, the recovery 
        points the queue's tail to the last traversed node.
        \item\label{liq-recovery-last-node-with-next-node-with-init-unset} If the traversal ends due to a node with an unset {\em initialized} flag, then let $P$ be its preceding node. The recovery sets $P.${\em next} to \nul{} and flushes it, and sets the tail to point to $P$.
    \end{enumerate}
\end{enumerate}
In all cases, the {\em pred} field of the last node (pointed to by the tail) is set to \nul{}. 
In addition, throughout the queue traversal, the addresses of all traversed nodes with a set {\em initialized} flag are recorded. All other nodes in the designated allocation areas are reclaimed.
For each reclaimed node with a set {\em initialized} flag, the recovery unsets the {\em initialized} flag and flushes it before retiring the node.  There could be at most two such nodes per thread: There is at most one such node (namely, a node which is not part of the queue but has a set {\em initialized} flag) which the thread has dequeued and placed in its local {\em nodeToPersistAndRetire} array, where the node awaits its persistence. In addition, there could be another such node -- a node that the thread was about to enqueue, if the thread were in the middle of an enqueue operation when the crash occurred; or alternately a node that the thread has just advanced the head from, if the thread were in the middle of a dequeue operation when the crash occurred.

If any flush were executed during the recovery, a single \sfence{} is placed in the end to ensure the completion of the executed flushes.

\section{\ouq{} Details}\label{ouq-details}
\Cref{fig:ouq-impl} contains the pseudocode of the \ouq{} algorithm, described in \Cref{ouq-details-description}. 
Note that the queue's global head and tail pointers point to \volatile{} nodes.
The \movnti{} instruction is a non-temporal store instruction that writes back data directly to the memory, bypassing the caches.

\begin{figure}[h]
\caption{\ouq{} implementation}\label{fig:ouq-impl}
\vspace{0.5\baselineskip}
\begin{minipage}{.49\textwidth}
\begin{lstlisting}
class Persistent
    Item* item
    int index
    bool linked
class Volatile
    Item* item
    int index
    atomic<Volatile*> next
    Persistent* persistentNode
\end{lstlisting}
\begin{lstlisting}
@\underline{Item* Dequeue()}@
    while (true)
        head = Head
        headNext = head->next@\label{ouq-read-head-next}@
        if (headNext == NULL)
            movnti(&localData[tid].headIndex, head->index)
            SFENCE@\label{ouq-persist-failing-deq}@
            return NULL
        if (CAS(&Head, head, headNext)@\label{ouq-deq-cas}@
            dequeuedItem = headNext->item
            movnti(&localData[tid].headIndex, headNext->index)
            SFENCE@\label{ouq-persist-successful-deq}@
            if (localData[tid].nodeToRetire)
                retire(localData[tid].nodeToRetire->persistentNode)
                retire(localData[tid].nodeToRetire)
            localData[tid].nodeToRetire = head@\label{ouq-save-node-to-reclaim}@
            return dequeuedItem
\end{lstlisting}
\end{minipage}\hfill
\begin{minipage}{.49\textwidth}
\begin{lstlisting}
@\underline{Enqueue(item)}@
    newNode = allocVolatile()@\label{ouq-alloc-node}@
    newNode->item = item
    newNode->next = NULL@\label{ouq-enq-set-next-null}@
    newNode->persistent = allocPersistent()
    newNode->persistent->item = item
    newNode->persistent->linked = false@\label{ouq-unset-linked}@
    while (true)
        tail = Tail
        if (tail->next == NULL)
            newNode->persistentNode->index = tail->index + 1
            newNode->index = newNode->persistentNode->index
            if (CAS(&tail->next, NULL, newNode))@\label{ouq-link}@
                newNode->persistentNode->linked = true@\label{ouq-set-linked}@
                FLUSH(newNode->persistentNode); SFENCE@\label{ouq-persist-enq}@
                CAS(&Tail, tail, newNode)@\label{ouq-advance-tail}@
                break
        CAS(&Tail, tail, tail->next)@\label{ouq-enq-assisting-cas-tail}@
\end{lstlisting}
\end{minipage}
\end{figure}
\section{\olq{} Details}\label{olq-details}
The pseudocode of the \olq{} algorithm appears in \Cref{fig:olq-impl-deq,fig:olq-impl-enq}. 
The queue's global head and tail pointers point to \volatile{} nodes.
{\em localData} is an array consisting of a cell per thread. Each thread may access its cell using its thread ID as an index. Each cell consists of the fields {\em headIndex} and {\em nodeToRetire} accessed in dequeues, and {\em lastEnqueues} (an array containing two cells, each composed of a pointer to a \persistent{} object and an index), {\em lastEnqueuesIndex} and {\em validBit} accessed in enqueues. 
{\em localData} array's cells do not share cache lines to avoid false sharing. 
In addition, for each cell, the {\em lastEnqueues} array and {\em headIndex} field, which are written using \movnti{} instructions, are kept in a cache line separate from the rest of the cell's fields.

Next, we describe \olq{}'s operations in detail.

\subsection{The Enqueue Operation}\label{olq-enq-description}
The enqueue operation first allocates a \volatile{} node denoted {\em newNode} from the memory manager and a matching \persistent{} node and initializes their data (\Crefrange{olq-alloc-node}{olq-init-end}).
Then, before attempting to link {\em newNode} to the last node, it sets the {\em pred} and {\em index} fields of both the \volatile{} and \persistent{} parts (\Crefrange{olq-set-pred}{olq-set-persistent-index}). The {\em index} field of the \persistent{} object serves as a stamp indicating to the recovery that the object's data is up-to-date: {\em index} is the last written field of the \persistent{} object, for ensuring that if this object is traversed during a recovery walk, and its {\em index} is identified as non-stale, then all the object's data is non-stale. This is due to 
\Cref{release-fence-adequate-between-writes-to-same-cache-line}, guaranteeing that the order of writing {\em index} after the other fields is preserved in NVRAM.

Next, the enqueue operation attempts to link {\em newNode} to the last \volatile{} node (\Cref{olq-link}), and on success it advances the queue's tail and ensures that the path of nodes leading from the head to {\em newNode->persistentNode} is flushed to the NVRAM (\Crefrange{olq-advance-tail}{olq-flushNotPersistedSuffix}). 
\begin{figure}[t]
\caption{\olq{} implementation -- Objects and Dequeue}\label{fig:olq-impl-deq}
\vspace{0.5\baselineskip}
\begin{minipage}{.49\textwidth}
\begin{lstlisting}
class Persistent
    Item* item
    Persistent* pred
    int index
class Volatile
    Item* item
    atomic<Volatile*> next
    atomic<Volatile*> pred
    int index
    Persistent* persistentNode
\end{lstlisting}
\begin{lstlisting}
@\underline{Item* Dequeue()}@
    while (true)
        head = Head@\label{olq-deq-read-head}@
        headNext = head->next@\label{olq-deq-read-headNext}@
        if (headNext == NULL)
            movnti(&localData[tid].headIndex, head->index)@\label{olq-failing-deq-writes-local-head}@
            SFENCE@\label{olq-failing-deq-persist-local-head}@
            return NULL@\label{olq-failing-deq-return-null}@
        if (CAS(&Head, head, headNext)@\label{olq-advance-head}@
            dequeuedItem = headNext->item
            movnti(&localData[tid].headIndex, headNext->index)@\label{olq-successful-deq-writes-local-head}@
            SFENCE@\label{olq-successful-deq-persist-head-fence}@
            headNext->pred = NULL@\label{olq-deq-set-pred-null}@
            if (localData[tid].nodeToRetire)@\label{olq-retire-start}@
                retire(localData[tid].nodeToRetire->persistentNode)
                retire(localData[tid].nodeToRetire)@\label{olq-retire-end}@
            localData[tid].nodeToRetire = head@\label{olq-save-node-to-reclaim}@
            return dequeuedItem
\end{lstlisting}
\end{minipage}\hfill
\end{figure}

It then records the address and index of the newly enqueued \persistent{} node in the thread's {\em lastEnqueues} array (\Cref{record-last-enqueue}). This array contains two cells per thread -- for keeping record of the thread's last and penultimate enqueued nodes. The thread writes alternately -- on each enqueue it writes to the cell with index {\em localData[tid].lastEnqueueIndex} and in the end flips its {\em lastEnqueueIndex} (in \Cref{olq-flip-index}).
The writes to {\em lastEnqueues} are performed using \movnti{} instructions (\Crefrange{olq-start-update-local-tails}{olq-end-update-local-tails}). In case a crash occurs after only one of the address and index was written to the memory, the subsequent recovery needs to identify that the cell's content is invalid and should be ignored. To this end, we place a valid bit in both the address and value (the least significant bit of the address and the most significant bit of the index). A {\em lastEnqueues} cell is considered valid only if the valid bits of its address and index match. After the writes, the value of {\em localData[tid].validBit} is flipped if {\em localData[tid].lastEnqueues=1} (\Cref{olq-flip-valid-bit}), so that the thread's following writes to its two {\em lastEnqueues} cells will be with the opposite valid bit value.

\begin{figure}[h]
\caption{\olq{} implementation -- Enqueue}\label{fig:olq-impl-enq}
\vspace{0.5\baselineskip}
\begin{minipage}{.49\textwidth}
\begin{lstlisting}
@\underline{FlushNotPersistedSuffix(notPersisted)}@
    while (true)
        pred = notPersisted->pred
        if (pred == NULL)
            break
        FLUSH(notPersisted->persistentNode)
        notPersisted = pred
@\underline{ZeroBit(value, bitIndex)}@
    return value & ~(1 << bitIndex)
@\underline{ApplyBit(value, bitIndex, bitValue)}@
    return ZeroBit(value, bitIndex) | (bitValue << bitIndex)
@\underline{RecordLastEnqueue(newNode)}@
    i = localData[tid].lastEnqueuesIndex
    movnti(&localData[tid].lastEnqueues[i].ptr, ApplyBit(newNode->persistentNode, 0, localData[tid].validBit))@\label{olq-start-update-local-tails}@
    movnti(&localData[tid].lastEnqueues[i].index, ApplyBit(newNode->index, sizeof(newNode->index)*8-1, localData[tid].validBit))@\label{olq-end-update-local-tails}@
    localData[tid].validBit ^= i // Flip valid bit if i=1@\label{olq-flip-valid-bit}@
    localData[tid].lastEnqueuesIndex ^= 1 // Flip index@\label{olq-flip-index}@
@\underline{Enqueue(item)}@
    newNode = allocVolatile()@\label{olq-alloc-node}@
    newNode->item = item
    newNode->next = NULL
    newNode->persistentNode = allocPersistent()
    newNode->persistentNode->item = item@\label{olq-init-end}@
    while (true)
        tail = Tail
        if (tail->next == NULL)
            newNode->pred = tail@\label{olq-set-pred}@
            newNode->index = tail->index + 1
            newNode->persistentNode->pred = tail->persistentNode
            newNode->persistentNode->index = newNode->index@\label{olq-set-persistent-index}@
            if (CAS(&tail->next, NULL, newNode))@\label{olq-link}@
                CAS(&Tail, tail, newNode)@\label{olq-advance-tail}@
                FlushNotPersistedSuffix(newNode)@\label{olq-flushNotPersistedSuffix}@
                RecordLastEnqueue(newNode)@\label{record-last-enqueue}@
                SFENCE@\label{olq-fence}@
                // All nodes up to newNode are persistent
                newNode->pred = NULL@\label{olq-enq-set-pred-null}@
                break
        CAS(&Tail, tail, tail->next)@\label{olq-enq-assisting-cas-tail}@
\end{lstlisting}
\end{minipage}
\end{figure}

Finally, the enqueue operation issues an \sfence{} (\Cref{olq-fence}) to ensure the completion of all executed flushes and \movnti{} instructions. In particular, all \persistent{} nodes succeeding the current head up to {\em newNode->persistentNode} are guaranteed to be persistent. To prevent future enqueues from redundantly flushing these nodes, the enqueuer then sets {\em newNode}'s backward link to \nul{} (\Cref{olq-enq-set-pred-null}). Thus, each enqueue operation that reaches {\em newNode} from now on, during its backward walk, would not need to traverse the preceding \persistent{} nodes. 

Like in the original \msq{}, a concurrent enqueue might prevent the enqueue's linking. In this case, the enqueuer tries to assist the obstructing enqueue and advance the tail to point to the node enqueued by that obstructing enqueue (\Cref{olq-enq-assisting-cas-tail}), before starting a new attempt to enqueue its own item.

\subsection{The Dequeue Operation}\label{olq-deq-description}
The dequeue operation attempts to extract the oldest item, placed in the node subsequent to the dummy node.
If the queue is empty when the dequeue operation takes effect, it returns \nul{}.
But before returning, the failing dequeue must ensure that previous dequeues that emptied the queue survive a crash. It does so by copying the head's index to its local head index and persisting it (\Crefrange{olq-failing-deq-writes-local-head}{olq-failing-deq-persist-local-head}). 
Each thread's local head index variable is placed in the thread's cell in the {\em localData} array.

If the queue is not empty, the dequeuer attempts to advance the head by one node (\Cref{olq-advance-head}), and on success -- returns the oldest item to the caller. On failure it retries the whole scheme.
Before returning, the dequeuer copies the new head's index to its local head index and persists it (\Crefrange{olq-successful-deq-writes-local-head}{olq-successful-deq-persist-head-fence}), to comply with durable linearizability, which requires that completed operations be linearized.  

A successful dequeue is responsible for reclaiming the dummy node recorded by the previous dequeue executed by the same thread.  
Before reclaiming, it must ensure that the node is unreachable by future operations. To make it unreachable by backward walks (of enqueue operations that will try to identify a non-persisted suffix), the dequeuer disconnects the backward link from the new dummy head to the previous one (\Cref{olq-deq-set-pred-null}).
It then returns the \persistent{} and \volatile{} objects of the previous dummy node to the memory manager (\Crefrange{olq-retire-start}{olq-retire-end}), and keeps a record of the current dummy node for its future reclamation (\Cref{olq-save-node-to-reclaim}).

\subsection{Recovery}
\label{olq-recovery-subsection}
The recovery procedure of \olq{} resurrects all nodes reachable through backward links from the abstract tail until the node succeeding the dummy head.
It then allocates matching \volatile{} objects and sets their forward links to form the linked list that constitutes the volatile queue.
This is implemented as follows. 

Let {\em headIndex} be the maximal index among the local head indices of all threads. The recovery does not modify these per-thread indices.
It sorts all per-thread {\em lastEnqueues}'s indices that are valid (namely, their valid bit value matches the valid bit value of the associated pointer), bigger than {\em headIndex} and have an associated non-\nul{} pointer from largest to smallest, and gathers them with their matching per-thread last enqueue pointers to a single list of potential tails. The recovery then attempts to start a backward walk from each potential pointer, one after another. For each attempted pointer, if the index in the \persistent{} object it points to is different from the associated index kept in the appropriate {\em lastEnqueues} cell, or if a nonconsecutive index is encountered during the backward walk from it to the \persistent{} object with index {\em headIndex+1} (each of these cases implies that the index of the inspected \persistent{} object is stale) -- the recovery moves on to try the next potential tail.

All \persistent{} objects in the designated allocation areas but the ones traversed in the last successful walk are reclaimed (if there was such a walk, otherwise the queue is empty and all \persistent{} objects are reclaimed). For each reclaimed node with an index bigger than {\em headIndex} (there could be at most one such node per thread -- for threads that were in the middle of enqueuing when the crash occurred), the recovery zeroes the node's index and flushes it before retiring the node.

In order to construct a linked list of \volatile{} objects, for each of the recovered \persistent{} objects, the recovery allocates a \volatile{} object and sets its \persistent{} pointer to the associated \persistent{} object. In addition, the {\em index} and {\em item} of each \volatile{} are copied from the associated \persistent{}. The {\em next} pointers of the \volatile{} objects are set according to the queue's order. The {\em pred} field of the last \volatile{} object is set to \nul{}. 
Dummy \volatile{} and \persistent{} objects are allocated too. Their {\em index} fields are set to {\em headIndex}. The \persistent{} pointer of the dummy \volatile{} object is pointed at the dummy \persistent{} object. The {\em next} pointer of the dummy \volatile{} is pointed at the recovered \volatile{} object with index {\em headIndex+1}, or set to \nul{} if an empty queue is recovered. 
The queue's head and tail pointers are pointed at the first and last \volatile{} objects in the linked list respectively.

For all threads that do not contain a valid record of the recovered tail in any of their {\em lastEnqueues} cells, these cells are zeroed using \movnti{} instructions. In addition, their {\em lastEnqueuesIndex} is set to 0, and their {\em validBit} is set to 1.
For a thread with a valid {\em lastEnqueues} cell referring to the recovered tail: Its other cell is zeroed using \movnti{} instructions. In addition, its {\em lastEnqueuesIndex} is set to the other cell's index, and the thread's {\em validBit} is set appropriately (so that the next write to the cell that refers to the recovered tail will be with a valid bit value opposite of its current one).

Finally, the recovery issues an \sfence{} to ensure the completion of all executed flushes and \movnti{} instructions.

\section{Lock-Freedom Proof}
\label{lock-freedom-section}
To prove lock-freedom in the presence of crashes, we need to prove that each time a thread executes an operation on the queue,
and there are no interrupting crash events since the operation's invocation,
some thread completes an operation on the queue within a finite number of steps.
Namely, it is sufficient to prove progress for crash-free intervals of execution. 
For each of the four described queue algorithms, the following holds: within $n$+1 loop iterations of a given running operation (assuming a crash-free long-enough interval of execution), where $n$ is the number of threads operating on the queue, some operation succeeds to perform a linearization point.

We complete the argument brought in \Cref{Lock-Freedom-sketch}. 
We start with noting that an obstructing \vlp{} of some operation does not cause another operation to branch backwards more than once: a dequeue obstructing another dequeue has advanced the head, so the interrupted dequeue will read a new value from the queue's head in its next iteration, and an enqueue interrupted by another enqueue ensures the tail is advanced before starting a new iteration.

Next, we explain why in case of a dequeue operation, $n$ iterations are sufficient to guarantee progress.
Let the examined running $op$ be a dequeue. It branches backwards each time another dequeue precedes it with advancing the head. If $op$ does not complete within $n$ iterations, some other thread must have advanced the head twice, in two different dequeue operations. This means it must have completed the first dequeue operation of the two, denoted {\em firstDeq}. Prior to completing {\em firstDeq}, the other thread has persisted the head. Thus, {\em firstDeq} is linearized. We still need to show that the linearization point occurs within the $n$ inspected iterations of $op$ and not prior to them, in order to show that $n$ iterations of a dequeue are enough to achieve progress. {\em firstDeq}'s linearization point occurs in $op$'s iteration in which {\em firstDeq} has failed $op$, because in this iteration $op$ read the queue's head, and then failed to advance it since the obstructing {\em firstDeq} has advanced it in between.

For an enqueue, $n$ iterations are not adequate to ensure progress.
Let the examined running $op$ be an enqueue. We analyze its execution since an iteration it started at moment $t$. $op$ branches backwards each time another enqueue precedes it with linking a node to the tail.
If a linearized enqueue fails $op$'s first linking attempt, it is not guaranteed that the linearization point of this enqueue occurs after $t$. But from $op$'s second iteration on, each enqueue that fails $op$ and is linearized -- is guaranteed to be linearized after $t$: it is linearized when it links its node to a previous node denoted {\em N}, after the tail is advanced to point to {\em N}, which happens after $op$ obtains the tail in the first inspected iteration (since its obtained value must point to a preceding node, to which another enqueue operation has linked a node).
Therefore, we do not look at the first $n$ iterations of $op$, but rather at the $n$ iterations starting with the second one. A similar argument to the one brought for a dequeue $op$ applies to these iterations: If $op$ does not complete within $n$+1 iterations, some other thread must have linked twice within iterations $2$ to $n$+1, in two different enqueue operations. This means it has completed the first enqueue of the two. Prior to returning from this enqueue, it has ensured the \pp{} of that enqueue. Thus, this enqueue is linearized. As explained before, its linearization point occurs after $t$, namely, within the $n$+1 inspected iterations of $op$.

\end{document}